\documentclass{article}

\usepackage[hmargin=0.5in,vmargin=2cm]{geometry}
\usepackage[margin=0.5cm,font=footnotesize,labelfont=bf,labelsep=colon]{caption}

\usepackage{authblk}

\usepackage{cite}
\usepackage{amsmath,amssymb,amsfonts}
\usepackage{mathtools}
\usepackage{xfrac}
\usepackage[cal=pxtx,scr=boondoxo]{mathalfa}
\usepackage{algorithmic}
\usepackage[extdef=true]{delimset}
\usepackage{graphicx}
\usepackage{textcomp}

\usepackage{xcolor}
\usepackage{soul}
\usepackage{sectsty}

\usepackage[T1]{fontenc}

\definecolor{NRLblue}{cmyk}{0.95,0.74,0.07,0.44}
\definecolor{NRLblue2}{cmyk}{0.90,0.62,0.08,0.01}

\partfont{\color{NRLblue2}}
\sectionfont{\color{NRLblue2}}
\subsectionfont{\color{NRLblue2}}
\subsubsectionfont{\color{NRLblue2}}

\newcommand*{\var}[1]{\mkern1mu\brk0{#1}}

\DeclareMathOperator{\Imag}{Im}
\DeclareMathOperator{\Real}{Re}

\newcommand*{\imj}{\mathscr{j}}
\newcommand*{\rnorm}{\varrho}

\newcommand*{\sigmach}{\bar{\sigma}_{C\!H}}

\begin{document}
\title{\color{NRLblue2} \textbf{\textsc{Modeling Practical Capacitive Structures for Electronic and Dielectric Property Extraction and Analysis}}}
\author[1,*]{James G. Champlain}
\author[1]{Vikrant J. Gokhale}
\author[1]{Matthew T. Hardy}
\author[1]{Eric N. Jin}
\affil[1]{U.S.\ Naval Research Laboratory, 4555 Overlook Ave. SW, Washington, DC 20375, USA}
\affil[*]{corresponding author: e-mail: james.g.champlain.civ@us.navy.mil}
\date{}
\maketitle

\begin{abstract}
A method for modeling the full steady-state and small-signal behavior of practical capacitive structures, such as metal-insulator-metal capacitors, diodes, and transistors, is presented.  Simple lumped element models fail to properly represent the behavior of most practical structures, resulting in incorrect electronic material property extraction and evaluation.  The methodology and model presented in this paper properly characterizes practical structures and can be employed to more accurately understand and evaluate experimental results, extract various electronic and dielectric properties of the constituent materials, and predict possible behavior of new structures and materials.  The impact of dielectric properties (e.g., real versus complex permittivity, conductance, Cole-Cole dielectric response) and geometry on observed behavior of practical structures, especially with regard to extracting correct property values, is discussed.
\end{abstract}

\section{Introduction}
\label{sec:introduction}
Capacitance--voltage (CV), capacitance-frequency (C$\! f$), and other more comprehensive impedance or admittance spectroscopy measurements are useful techniques in evaluating the electronic and dielectric behavior and properties of semiconductor and dielectric materials, such as relative permittivity, conductivity, and dissipation factor or loss tangent \cite{Lang1987,Macdonald1992,Paszkiewicz2001,Shealy2008,Paszkiewicz2013,Donahue2013,Kohler2015}.  Often in the analysis of these measurements on practical electronic devices, the simplest models are employed, and for the extraction of the most basic of information, such models may be sufficient \cite{Wu2017,Casamento2022}.  However, in cases where the device or material being modeled has more complex behavior than conventionally ``ideal'' capacitors (i.e., cannot be approximated by simple one-dimensional geometries with materials consisting of real-valued, frequency independent permittivities), a more comprehensive approach is required that is capable of capturing the more complex dielectric behavior and responses of real-world devices and materials.

In this paper, a method for modeling the full steady-state and small-signal behavior of practical capacitive structures (``capacitors''), such as metal-insulator-metal (MIM) capacitors, Schottky or pn diodes, and field-effect transistors (FETs), is presented.  The subsequent methodology and model can then be used to understand and evaluate experimental results, to extract various electronic and dielectric properties of the constituent materials, and to predict possible behavior of new structures and materials.  Examples of practical concern will be presented, with comparisons between common, simplistic interpretations of measured data and a model developed following the methodology laid out in this paper shown.  The impact of dielectric properties (e.g., ideal real versus complex permittivity, dielectric relaxation, conductivity) and geometry on observed behavior is discussed.

Lastly, the methodology presented in this paper is applied to two practical examples to demonstrate its validity: a MIM capacitor using epitaxially grown ScAlN and a high-electron mobility transistor (HEMT) incorporating epitaxially grown barium titanate (BTO) passivation.  Both ScAlN and BTO are emerging thin film materials that have presented properties of significant interest and are currently the focus of a great deal of contemporary research \cite{Casamento2022,Giribaldi2022,Wang2020,Pradhan2024,Xia2019,Rahman2021}.  As a result, they are expected to play a large role in future electronic devices and are therefore excellent candidates to demonstrate the wide-ranging applicability of the methodology presented in this paper.

\section{Model Development}
\begin{figure}[!htb]
	\centerline{\includegraphics[width=7.05cm]{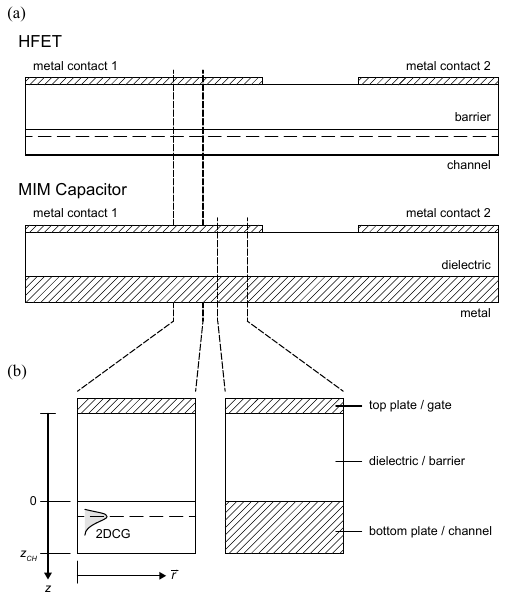}}
	\caption{(a) Cross-section of example ``capacitors'', a heterostructure field-effect transistor (HFET) and metal-insulator-metal (MIM) capacitor.  (b) Highlighted cross-sections of the core structure of these capacitors, indicating the top plate or gate, the bottom plate or channel, and the dielectric or barrier.}
	\label{fig:generic_cap}
\end{figure}

The different possible capacitors that could be evaluated via various impedance spectroscopy techniques span a large range of materials and device structures; however, as will be shown, they also share a number of similarities that allow for a single general, yet comprehensive analysis to be applied to them all.  Fig.~\ref{fig:generic_cap} shows two possible examples of capacitors that might be studied using impedance spectroscopy or similar methods: a heterostructure FET (HFET) and an MIM capacitor.  Both consist of a material stack, two metallic contacts on the surface, and a conductive layer underneath.  Irrespective of the specifics, the core structures of these capacitors share a similar construction (Fig.~\ref{fig:generic_cap}b) consisting of a top ``plate,'' a bottom ``plate,'' and a ``dielectric'' between.

It is assumed the top plate is a highly conductive metal and can be treated as a lossless equipotential.  The bottom plate could be any number of possible conductive ``channels'', including but not limited to a two-dimensional carrier gas (e.g., the channel of a FET), a lossy metal (e.g., the bottom plate of an MIM capacitor), or a doped bulk semiconductor layer (e.g., one side of a pn junction or the body of a MOS capacitor).  It is critical to note that what constitutes the bottom plate is not necessarily defined by a specific material layer or interface, but instead by the existence of a conductive pathway.  The dielectric of the capacitor between the plates could be any material or combination of materials.  And like the bottom plate, the dielectric is not defined by a specific material or layer structure but by the existence of a capacitive element, including but not limited to a conventional insulating dielectric, a layered semiconductor structure that creates an electronic barrier, the depletion layer of a conventionally doped semiconductor, and/or the quantum capacitance of a two-dimensional carrier gas.  Furthermore, it need not be assumed that the dielectric is perfectly insulating, though the more conductive the layer is, the more complicated the analysis becomes.  Borrowing from FET terminology, for the remainder of this paper the top plate will be referred to as the ``gate,'' the bottom plate as the ``channel,'' and the dielectric of the capacitor between as the ``barrier,'' without any inferences regarding charge behavior or transport. 

The analysis of the capacitor begins with the continuity equation:
\begin{equation}\label{eq:cont_eq}
	\nabla \cdot \vec{\jmath}_F \var{ \vec{R},t } = - \frac{\partial}{\partial t} \rho \var{ \vec{R},t }
\end{equation}
where $\vec{\jmath}_F \var{ \vec{R},t }$ is the (free) conduction current density (A/cm$^2$) and $\rho \var{ \vec{R},t }$ is the charge density (C/cm$^3$), both at position $\vec{R}$ and time $t$.  Gauss' law relates the charge density to the electric displacement field (C/cm), $\vec{D} \var{ \vec{R},t }$:
\begin{equation}
	\rho \var{ \vec{R},t } = \nabla \cdot \vec{D} \var{ \vec{R},t } \, ,
\end{equation}
which when inserted into the continuity equation leads to
\begin{equation}
		\nabla \cdot \vec{\jmath}_F \var{ \vec{R},t } = - \frac{\partial}{\partial t} \brk*{ \nabla \cdot \vec{D} \var{ \vec{R},t } } \\
		= - \nabla \cdot \brk*{ \frac{\partial}{\partial t} \vec{D} \var{ \vec{R},t } } \\
		= - \nabla \cdot \vec{\jmath}_D \var{ \vec{R},t }
\end{equation}
where $\vec{\jmath}_D \var{ \vec{R},t } = \sfrac{\partial}{\partial t} \brk{ \vec{D} \var{ \vec{R},t } }$ is the electric displacement current density (A/cm$^2$) associated with a time-varying electric displacement field.  The continuity equation of \eqref{eq:cont_eq}, initially between conduction current and charge density, can now be understood as a continuity equation between conduction and displacement currents:
\begin{equation}\label{eq:j_total}
	\nabla \cdot \brk*{ \vec{\jmath}_F \var{\vec{R},t} + \vec{\jmath}_D \var{\vec{R},t} } = \nabla \cdot \vec{\jmath}_T \var{\vec{R},t} = 0
\end{equation}
where $\vec{\jmath}_T \var{\vec{R},t} = \vec{\jmath}_F \var{\vec{R},t} + \vec{\jmath}_D \var{\vec{R},t}$ is the ``total'' (conduction plus displacement) current density.  In effect, \eqref{eq:j_total} is equivalent to Kirchhoff’s first circuit law (i.e., Kirchhoff's current law) in three-dimensional space.  Fundamentally there is no difference between this expression and the continuity equation of \eqref{eq:cont_eq}, \eqref{eq:j_total} simply allows for an alternate expression of the underlying physics useful for this analysis.

The remainder of the analysis in this paper will be examining the capacitor under small-signal conditions, where the small-signal approximation applies.  Under the small-signal approximation, the full, time-dependent expression of a term can be expressed as a sum of its steady-state (DC) and small-signal (ac) components.  For the sake of clarity, the following notation will be used to differentiate these terms when possible.  Using the total current density as an example:
\begin{equation}
	\vec{\jmath}_T \var{\vec{R},t} = \vec{J}_T \var{\vec{R}} + \vec{\jmath}_t \var{\vec{R},t}
\end{equation}
where $\vec{\jmath}_T \var{\vec{R},t}$ is the full (steady-state plus small-signal) total current density, denoted by lowercase notation and uppercase subscript; $\vec{J}_T \var{\vec{R}}$ is the steady-state total current density, denoted by uppercase notation and uppercase subscript; and $\vec{\jmath}_t \var{\vec{R},t}$ is the small-signal total current density, denoted by lowercase notation and lowercase subscript.  Where such notation is not possible, alternative notation will be defined clearly.  Furthermore, phasor notation will be used for the small-signal terms and associated expressions.  For example, continuing with the small-signal total current density:
\begin{subequations}
	\begin{equation}
		\vec{\jmath}_t \var{\vec{R},t} \equiv \vec{\mathbb{J}}_t \var{\vec{R}} e^{\imj \omega t}
	\end{equation}
	where
	\begin{equation}\label{eq:phasor}
		\vec{\mathbb{J}}_t \var{\vec{R}} = \brk*{ \mathbb{J}_{t,x_1} \var{\vec{R}} e^{ \imj \theta_{\mathbb{J}_t,x_1} \var{\vec{R}} } , \\
		\mathbb{J}_{t,x_2} \var{\vec{R}} e^{ \imj \theta_{\mathbb{J}_t,x_2} \var{\vec{R}} } , \\
		\mathbb{J}_{t,x_3} \var{\vec{R}} e^{ \imj \theta_{\mathbb{J}_t,x_3} \var{\vec{R}} } }
	\end{equation}
\end{subequations}
is the phasor of the small-signal total current density, carrying only information about the magnitude (i.e., $\mathbb{J}_{t,x_n} \var{\vec{R}}$, $n=1,2,3$) and phase (i.e., $\theta_{\mathbb{J}_t,x_n} \var{\vec{R}}$, $n = 1,2,3$) of the term; $\imj = \sqrt{-1}$ is the imaginary unit; and $\omega = 2 \pi f$ is the fundamental frequency of the small-signal stimulus.
In the example here, the current density is a vector quantity and therefore its phasor in \eqref{eq:phasor} is also a vector quantity, presented as a vector triplet in an arbitrary 3D-space of $\vec{R} \equiv \brk*{ x_1, x_2 , x_3 }$.
The use of phasor notation simplifies the expression of any derivatives with respect to time and subsequently any impedances and admittances.

Due to the planar nature of the capacitors considered here, it will be useful within this analysis to decompose the three-dimensional position vector, $\vec{R}$, into a lateral position vector, $\vec{r}$, that lies within the plane of the gate or channel and a vertical position vector, $\vec{z} = z \, \hat{z}$, that is orthogonal to the lateral position vector: $\vec{R} = \vec{r} + \vec{z} \equiv \brk{\vec{r},z}$ (Fig.~\ref{fig:generic_cap}b).  Subsequently, any vector quantity, such as the total current density, can be similarly decomposed:
\begin{equation}
	\vec{\jmath}_T \var{\vec{R},t} = \vec{\jmath}_{T,r} \var{\vec{r},z,t} + j_{T,z} \var{\vec{r},z,t} \hat{z}
\end{equation}
where $\vec{\jmath}_{T,r} \var{\vec{r},z,t}$ and $j_{T,z} \var{\vec{r},z,t} \hat{z}$ are the lateral and vertical components of the total current density, respectively.  In the preceding and subsequent expressions, $\hat{z}$ is the unit vector in the vertical direction, pointing from gate to channel.

Applying \eqref{eq:j_total} and expanding the current density into its lateral and vertical components results in
\begin{equation}\label{eq:cont_eq_rz}
	\nabla_r \cdot \vec{\jmath}_{T,r} \var{ \vec{r},z,t } + \frac{\partial}{\partial z} j_{T,z} \var{\vec{r},z,t} = 0
\end{equation}
where $\nabla_r$ is the two-dimensional Del operator in the lateral plane: $\nabla = \nabla_r + \brk{\partial / \partial z} \hat{z}$.  Integration of \eqref{eq:cont_eq_rz} vertically across the channel ($z = \intv[c]{0}{z_{C\!H}}$) leads directly to
\begin{equation}\label{eq:cont_eq_ch}
	\nabla_r \cdot \vec{k}_{C\!H} \var{\vec{r},t} - j_{GC\!H} \var{\vec{r},t} = 0
\end{equation}
where
\begin{equation}
	\vec{k}_{C\!H} \var{\vec{r},t} = \int_0^{z_{C\!H}} \! \vec{\jmath}_{T,r} \var{\vec{r},z,t} \, dz
\end{equation}
is the total lateral current sheet density (A/cm) in the channel and
\begin{equation}
		j_{GC\! H} \var{\vec{r},t} = - \int_0^{z_{C\!H}} \! dj_{T,z} \var{\vec{r},z,t} \\
		= - \brk*{ j_{T,z} \var{\vec{r},z_{C\!H},t} - j_{T,z} \var{\vec{r},0,t} } \\
		= j_{T,z} \var{\vec{r},0,t}
\end{equation}
is the total vertical current density entering the channel from the gate (i.e., directed from gate to channel), where it has been assumed that no current exits out the back of the channel: $j_{T,z} \var{\vec{r},z_{C\! H},t} = 0$.  Equation \eqref{eq:cont_eq_ch} now represents a continuity equation for current \emph{within} the channel at any point laterally in the plane of the channel.

It is assumed that the conductivity (S/cm) of the channel, $\sigma \var{\vec{r},z,t}$, is sufficiently high that any electric displacement current in the channel can be ignored relative to the conduction current, i.e., $\vec{\jmath}_D \var{\vec{r},z,t} \ll \vec{\jmath}_F \var{\vec{r},z,t}$ for $z = \intv[c]{0}{z_{C\!H}}$.  The total lateral channel current sheet density can then be written as
\begin{equation}
	\vec{k}_{C\!H} \var{\vec{r},t} = \sigma_{C\!H} \var{\vec{r},t} \vec{E}_r \var{\vec{r},t}
\end{equation}
where $\sigma_{C\!H} \var{\vec{r},t}$ is the sheet conductance (S or S$\cdot\square$) of the channel given by
\begin{equation}
	\sigma_{C\!H} \var{\vec{r},t} = \int_0^{z_{C\!H}} \! \sigma \var{\vec{r},z,t} \, dz
\end{equation}
and $\vec{E}_r \var{\vec{r},t}$ is the lateral electric field (V/cm) in the channel, assumed uniform vertically across the conductive channel.  Given the assumptions of a sufficiently conductive channel and no current exiting the back of the channel, there should be very little to no potential drop vertically in the conductive channel, making this a valid assumption.  The lateral electric field in the channel can be expressed in terms of a local potential proportional to the local Fermi level:
\begin{equation}
		\vec{E}_r \var{\vec{r},t} = - \nabla_r v_{V\!\!AC} \var{\vec{r},t} = - \nabla_r \brk*{ - \frac{\mathcal{E}_{V\!\!AC} \var{\vec{r},t}}{q} } \\
		= - \nabla_r \brk*{ - \frac{\mathcal{E}_F \var{\vec{r},t} + \phi}{q} } \\
		= - \nabla_r v_{C\!H} \var{\vec{r},t}
\end{equation}
where $v_{V\!\!AC} \var{\vec{r},t}$ and $\mathcal{E}_{V\!\!AC} \var{\vec{r},t} = -q v_{V\!\!AC} \var{\vec{r},t}$ are the vacuum level potential (V) and potential energy (eV), respectively; $q = 1.602\times10^{-19}$~C is the elementary electronic charge; $\mathcal{E}_F \var{\vec{r},t}$ is the Fermi level energy (eV); $\phi = \mathcal{E}_{V\!\!AC} \var{\vec{r},t} - \mathcal{E}_F \var{\vec{r},t}$ is the work function (eV) of the channel material, assumed constant; and $v_{C\!H} \var{\vec{r},t} = - \mathcal{E}_F \var{\vec{r},t} / q$ is the local potential (V) in the channel.  The total lateral current sheet density in the channel can now be expressed in terms of this local potential:
\begin{equation}
	\vec{k}_{C\!H} \var{\vec{r},t} = - \sigma_{C\!H} \var{\vec{r},t} \nabla_r v_{C\!H} \var{\vec{r},t} \, .
\end{equation}
Additionally, it is useful to introduce the gate potential, $v_G \var{t}$, into the expression for the total lateral current sheet density.  Because the gate has been assumed to be an equipotential, the gate potential can be inserted into the gradient without introducing any error:
\begin{equation}\label{eq:kCH_vGCH}
		\vec{k}_{C\!H} \var{\vec{r},t} = - \sigma_{C\!H} \var{\vec{r},t} \nabla_r v_{C\!H} \var{\vec{r},t} \\
		= \sigma_{C\!H} \var{\vec{r},t} \nabla_r \brk*{ v_G \var{t} - v_{C\!H} \var{\vec{r},t} } \\
		= \sigma_{C\!H} \var{\vec{r},t} \nabla_r v_{GC\!H} \var{\vec{r},t}
\end{equation}
where $v_{GC\!H} \var{\vec{r},t} = v_G \var{t} - v_{C\!H} \var{\vec{r},t}$ is gate-channel potential difference.

Compared to the total lateral sheet current density, the total vertical current density entering the channel, $j_{GC\!H} \var{\vec{r},t}$, can be much more difficult to define due to the large number of possible ``barriers'' (i.e., material combinations and electronic structures) and associated charge transport mechanisms constituting any possible capacitor.  However, for a given steady-state and small-signal condition, it can be generally understood that the barrier can be represented by some form of conductance, associated with actual charge conduction vertically through the capacitor (barrier), and some form of capacitance, associated with electric displacement current and time-varying electric fields through the capacitor (barrier).  In other words, a general admittance can be attributed to the barrier:
\begin{subequations}
	\begin{equation}\label{eq:jGCH_def}
		j_{GC\!H} \var{\vec{r},t} = Y_B \var{\omega,\vec{r}} v_{GC\!H} \var{\vec{r},t}
	\end{equation}
	\begin{equation}\label{eq:YB_def}
		Y_B \var{\omega,\vec{r}} = G_B \var{\omega,\vec{r}} + \imj \omega C_B \var{\omega,\vec{r}}
	\end{equation}
\end{subequations}
where $Y_B \var{\omega,\vec{r}}$ is this barrier admittance (S/cm$^2$), comprised of the parallel combination of a conductance, $G_B \var{\omega,\vec{r}}$ (S/cm$^2$), and capacitance, $C_B \var{\omega,\vec{r}}$ (F/cm$^2$).  Although the admittance is presented as an seemingly simple parallel combination of conductance and capacitance, it should be understood that this is just a general representation of possibly more complex charge transport and dynamics and associated conductance/resistance-capacitance networks.  As a result, both the conductance and capacitance presented here could have some form of frequency dependence themselves, which is why the frequency has been explicitly noted in the functional dependence of these terms.

Inserting \eqref{eq:kCH_vGCH} and \eqref{eq:jGCH_def} into the channel continuity equation of \eqref{eq:cont_eq_ch} leads to
\begin{equation}\label{eq:cont_eq_diff_eq}
	\hat{\nabla}_r \cdot \brk*{ \sigma_{C\!H} \var{\vec{r},t} \hat{\nabla}_r v_{GC\!H} \var{\vec{r},t} } \\
	- Y_B \var{\omega,\vec{r}} \lambda^2 v_{GC\!H} \var{\vec{r},t} = 0
\end{equation}
and it can be seen that the continuity equation is beginning to take on the form of a (multi-dimensional) second-order differential equation, which could be interpreted as a sort of characteristic equation for the gate-channel potential, $v_{GC\!H} \var{\vec{r},t}$.  Note, \eqref{eq:cont_eq_diff_eq} has been normalized to a characteristic length within the lateral plane of the capacitor, $\lambda$, with $\hat{\nabla}_r \equiv \lambda \nabla_r$.  The specific value for this characteristic length is not overly critical and will be dependent upon the specific geometry of the capacitor, however it is assumed to be a constant value for a given capacitor.

The terms in \eqref{eq:cont_eq_diff_eq} can be expanded into their steady-state and small-signal components and an associated set of steady-state and small-signal equations can be written:
\begin{subequations}\label{eq:cont_eq_split}
	\begin{equation}\label{eq:cont_eq_steady_state}
		\hat{\nabla}_r \cdot \brk*{ \sigmach \var{\vec{r}} \hat{\nabla}_r V_{GC\!H} \var{\vec{r}} } \\
		- G_B \var{0,\vec{r}} \lambda^2 V_{GC\!H} \var{\vec{r}} = 0
	\end{equation}
	\begin{equation}\label{eq:cont_eq_small_sig}
		\hat{\nabla}_r \cdot \brk*{ \sigmach \var{\vec{r}} \hat{\nabla}_r v_{gch} \var{\vec{r},t} + \sigma_{ch} \var{\vec{r},t} \hat{\nabla}_r V_{GC\!H} \var{\vec{r}} } \\
		- Y_B \var{\omega,\vec{r}} \lambda^2 v_{gch} \var{\vec{r},t} = 0
	\end{equation}
\end{subequations}
where $\sigmach \var{\vec{r}}$ and $\sigma_{ch} \var{\vec{r},t}$ are the steady-state and small-signal components of the channel sheet conductance, $\sigma_{C\!H} \var{\vec{r},t} = \sigmach \var{\vec{r}} + \sigma_{ch} \var{\vec{r},t}$, respectively, and $G_B \var{0,\vec{r}} = Y_B \var{0,\vec{r}}$ is the steady-state conductance through the barrier.  Equation \eqref{eq:cont_eq_steady_state} describes the behavior of the capacitor at steady-state while \eqref{eq:cont_eq_small_sig} describes the small-signal behavior at the fundamental frequency, $\omega = 2 \pi f$.  There is strictly a third equation that results from this expansion:
\begin{equation*}
	\hat{\nabla}_r \cdot \brk*{ \sigma_{ch} \var{\vec{r},t} \hat{\nabla}_r v_{gch} \var{\vec{r},t} } = 0 \, .
\end{equation*}
However, this corresponds to behavior at the first harmonic ($2 \omega$), which under small-signal conditions and the small-signal approximation is considered negligible relative to the steady-state and fundamental small-signal responses and can therefore be ignored.

In any analysis where a solution is expected, it can be assumed that the functional dependencies of the steady-state sheet conductance, $\sigmach \var{\vec{r}}$, and barrier admittance, $Y_B \var{\omega,\vec{r}}$, on the steady-state bias conditions (e.g., the steady-state gate-channel potential) are known from the material and electronic structure of the capacitor.  As a result, the one unknown remaining in the set of expressions in \eqref{eq:cont_eq_split}, beyond the steady-state and small-signal gate-channel potentials to be solved for, is the small-signal channel sheet conductance, $\sigma_{ch}\var{\vec{r},t}$.  Therefore, in order to solve for the steady-state and small-signal gate-channel potentials, the small-signal channel sheet conductance must be defined in some way.

One method to defining the steady-state and small-signal components of a term is through a first-order Taylor expansion of that term about an appropriate steady-state condition.  For most capacitors to be evaluated, the channel sheet conductance will have an explicit dependence on the gate-channel potential across the capacitor.  Therefore, it is appropriate to expand the channel sheet conductance with respect to the steady-state gate-channel potential:
\begin{equation}
	\sigma_{C\!H} \var{\vec{r},t} = \eval2{ \sigma_{C\!H} \var{\vec{r},t} }_{V_{GC\!H} \var{\vec{r}}} \\
	+ \brk*{ v_{GC\!H} \var{\vec{r},t} - V_{GC\!H} \var{\vec{r},t} } \eval4{ \frac{\partial \sigma_{C\!H} \var{\vec{r},t}}{\partial v_{GC\!H} \var{\vec{r},t}} }_{V_{GC\!H} \var{\vec{r}}} \, ,
\end{equation}
which when equated to the small-signal approximation representation of the channel sheet conductance, $\sigma_{C\!H} \var{\vec{r},t} = \sigmach \var{\vec{r}} + \sigma_{ch} \var{\vec{r},t}$, leads to expressions for the steady-state and small-signal components of the channel sheet conductance:
\begin{subequations}\label{eq:cond_split}
	\begin{equation}
		\sigmach \var{\vec{r}} = \eval2{ \sigma_{C\!H} \var{\vec{r},t} }_{V_{GC\!H} \var{\vec{r}}}
	\end{equation}
	\begin{equation}\label{eq:cond_small_sig}
			\sigma_{ch} = \brk*{ v_{GC\!H} \var{\vec{r},t} - V_{GC\!H} \var{\vec{r},t} } \eval4{ \frac{\partial \sigma_{C\!H} \var{\vec{r},t}}{\partial v_{GC\!H} \var{\vec{r},t}} }_{V_{GC\!H} \var{\vec{r}}} \\
			= v_{gch} \var{\vec{r},t} \frac{\partial \sigmach \var{\vec{r}}}{\partial V_{GC\!H} \var{\vec{r}}} \, .
	\end{equation}
\end{subequations}
Insertion of the small-signal channel sheet conductance of \eqref{eq:cond_small_sig} into \eqref{eq:cont_eq_small_sig} results in
\begin{equation*}
	\hat{\nabla}_r \cdot \brk*{ \sigmach \var{\vec{r}} \hat{\nabla}_r v_{gch} \var{\vec{r},t} + v_{gch} \var{\vec{r},t} \hat{\nabla}_r \sigmach \var{\vec{r}} } \\
	- Y_B \var{\omega,\vec{r}} \lambda^2 v_{gch} \var{\vec{r},t} = 0 \, .
\end{equation*}
The expression inside the divergence operation can simply be interpreted as the result of the product rule applied to the gradient of the ``compound'' function $\sigmach \var{\vec{r}} v_{gch} \var{\vec{r},t}$.  As a result, the set of expressions in \eqref{eq:cont_eq_split} can now be expressed as
\begin{subequations}\label{eq:char_eq_split}
	\begin{equation}\label{eq:char_eq_steady_state}
		\hat{\nabla}_r \cdot \brk*{ \sigmach \var{\vec{r}} \hat{\nabla}_r V_{GC\!H} \var{\vec{r}} } \\
		- G_B \var{0,\vec{r}} \lambda^2 V_{GC\!H} \var{\vec{r}} = 0
	\end{equation}
	\begin{equation}\label{eq:char_eq_small_sig}
		\hat{\nabla}_r^2 \brk*{ \sigmach \var{\vec{r}} v_{gch} \var{\vec{r},t} } \\
		- \frac{Y_B \var{\omega,\vec{r}} \lambda^2}{\sigmach \var{\vec{r}}} \sigmach \var{\vec{r}} v_{gch} \var{\vec{r},t} = 0
	\end{equation}
\end{subequations}
where $\hat{\nabla}_r^2 \brk*{ \cdots } \equiv \hat{\nabla}_r \cdot \hat{\nabla}_r \brk*{ \cdots}$, following conventional notation.
As previously, the set of expressions in \eqref{eq:char_eq_split} capture the behavior of the capacitor under steady-state and small-signal conditions, and with knowledge of the physical behavior of the capacitor, they provide a means to evaluate the steady-state and small-signal gate-channel potential profiles ($V_{GC\!H} \var{\vec{r}}$ and $v_{gch} \var{\vec{r},t}$, respectively) within the capacitor.

Similar to \eqref{eq:char_eq_split}, the total lateral current sheet density in the channel in \eqref{eq:kCH_vGCH} can likewise be expanded into its steady-state and small-signal components, which will be useful in the evaluation of the capacitor:
\begin{subequations}\label{eq:kCH_def}
	\begin{equation}\label{eq:KCH_def}
		\vec{K}_{C\!H} \var{\vec{r}} = \sigmach \var{\vec{r}} \nabla_r V_{GC\!H} \var{\vec{r}}
	\end{equation}
	\begin{equation}\label{eq:kch_def}
			\vec{k}_{ch} \var{\vec{r},t} = \sigmach \var{\vec{r}} \nabla_r v_{gch} \var{\vec{r},t} + \sigma_{ch} \nabla_r V_{GC\!H} \var{\vec{r}}  \\
			= \nabla_r \brk*{ \sigmach \var{\vec{r}} v_{gch} \var{\vec{r},t} }
	\end{equation}
\end{subequations}
where in \eqref{eq:kch_def}, the result of \eqref{eq:cond_small_sig} is employed.  Like the previous set of equations, there is a third equation resulting from the expansion corresponding to current at the first harmonic, which is also considered negligible and can be ignored: $0 = \sigma_{ch} \var{\vec{r},t} \nabla_r v_{gch} \var{\vec{r},t}$.  With \eqref{eq:char_eq_split} and \eqref{eq:kCH_def} and information on the capacitor geometry and corresponding boundary conditions, the impedance or admittance of any capacitor can be evaluated.

\subsection{Cases of Practical Interest: The Metallic Channel and The Perfectly Insulating Barrier}

While a completely general solution is not possible, two special cases of immense practical interest emerge if the coefficient in \eqref{eq:char_eq_small_sig}, $Y_B \var{\omega,\vec{r}} \lambda^2 / \sigmach \var{\vec{r}}$, can be taken as constant with respect to lateral position within the capacitor: the metallic channel and the perfectly insulating barrier.

For capacitors with a channel that is metallic, the channel sheet conductance ($\sigmach \var{\vec{r}}$) is by definition very high and bias independent.  For the vast majority of cases involving these metallic channels, the barrier is commonly composed of non-ferroelectric dielectric materials and the associated barrier admittance ($Y_B \var{\omega,\vec{r}}$) can also be assumed bias independent, leading to both the channel sheet conductance and barrier admittance being constant with respect to lateral position.  Therefore, the coefficient in \eqref{eq:char_eq_small_sig} is constant and the equation can directly be solved for $\sigmach \var{\vec{r}} v_{gch} \var{\vec{r},t}$; and consequently $v_{gch} \var{\vec{r},t}$, given $\sigmach \var{\vec{r}}$ is known.  With a solution for $\sigmach \var{\vec{r}} v_{gch} \var{\vec{r},t}$, the small-signal total lateral current sheet density in the channel can be found via \eqref{eq:kch_def}, and with expressions for both the small-signal voltage and current within the capacitor, an expression for the admittance or impedance of the capacitor is directly found.  While ultimately dependent upon the specific nature of the capacitor structure, this special case of a metallic channel is outstanding in that the model can provide solutions for \eqref{eq:char_eq_split} and \eqref{eq:kCH_def} and the admittance/impedance of the capacitor even for fairly leaky barriers (i.e., high barrier conductance, $G_B \var{\omega,\vec{r}} \gg 0$), as this does not impact the constancy of the channel sheet conductance or barrier admittance with respect to lateral position.

For the case of a perfectly insulating barrier (i.e., $G_B \var{0,\vec{r}} = 0$), the analysis becomes a bit more involved.  For a perfectly insulating barrier, \eqref{eq:char_eq_steady_state} reduces to
\begin{equation}\label{eq:char_eq_steady_state_Perfect}
	\nabla_r \cdot \brk*{ \sigmach \var{\vec{r}} \nabla_r V_{GC\!H} \var{\vec{r}} } = 0
\end{equation}
where the normalization with respect to the characteristic length, $\lambda$, has been removed.  The term inside the divergence is readily identified as the total steady-state lateral current sheet density in the channel, $\vec{K}_{C\!H} \var{\vec{r}}$, from \eqref{eq:KCH_def}.  It can be directly shown that \eqref{eq:char_eq_steady_state_Perfect} requires the total steady-state lateral current in the channel, $\vec{I}_{C\!H} \var{\vec{r}}$ (A), to be constant with respect to lateral position, $\vec{r}$ (see Appendix~\ref{app:const_channel_current}).  Leveraging this result, the general procedure to finding the impedance or admittance of a capacitor with a perfectly insulating barrier is as follows.

Regardless of the specific geometry, without any sources or sinks of charge, the current must go to zero at some point within the channel of the capacitor, for example at the center of an inner contact or the outermost edge of an outer contact.  Therefore, the steady-state current, and subsequently current density, must be zero at all points within the channel of the capacitor as a result of \eqref{eq:char_eq_steady_state_Perfect}.  Consequently, this requires the steady-state gate-channel potential ($V_{GC\!H} \var{\vec{r}}$) to be constant across the capacitor, which subsequently, assuming a uniform channel (i.e., uniform material composition and structure), results in a constant channel sheet conductance ($\sigmach \var{\vec{r}}$) and barrier admittance ($Y_B \var{\omega,\vec{r}}$) across the lateral expanse of the capacitor.  With both the channel sheet conductance and barrier admittance constant with respect to lateral position, \eqref{eq:char_eq_small_sig} is now directly solvable for the compound function $\sigmach \var{\vec{r}} v_{gch} \var{\vec{r},t}$ and therefore $v_{gch} \var{\vec{r},t}$, given $\sigmach \var{\vec{r}}$ is assumed known from the previously evaluated steady-state condition.  With a solution for $\sigmach \var{\vec{r}} v_{gch} \var{\vec{r},t}$, the small-signal total lateral current sheet density in the channel can again be found via \eqref{eq:kch_def}.  And finally, with expressions for both the small-signal voltage and current within the capacitor, an expression for the admittance or impedance of the capacitor is again directly found.  In practice, most capacitors of interest do not have a perfectly insulating barrier.  However, such cases where the barrier is still much, much less conductive than the channel and $G_B \var{0,\vec{r}} \lambda^2 \ll \sigmach \var{\vec{r}}$, the barrier can be considered nearly insulating and be approximated by the perfectly insulating case given by \eqref{eq:char_eq_steady_state_Perfect}, thereby allowing the same method of solution to be applied.

For cases where such approximations are not possible, alternative solution methods, which lie outside the scope of this paper, must be employed.

\section{Capacitors with Circular Geometry}
The expressions in \eqref{eq:char_eq_split} and \eqref{eq:kCH_def} can be applied to any geometry that is quasi-planar.  The vast majority of structures often employed in general material and device examinations are circular in nature (e.g., CV dots, circular diodes, mercury probes), therefore an exemplar circular capacitor will be examined here.  The result of a similar analysis on a rectilinear geometry with a high aspect ratio is presented in Appendix~\ref{app:rectilinear_cap}.

Fig.~\ref{fig:circular_cap} shows the structure of a generic capacitor of circular geometry, consisting of an inner circular contact, an outer annular contact, and a gap between.
\begin{figure}[!htb]
	\centerline{\includegraphics[width=7.0cm]{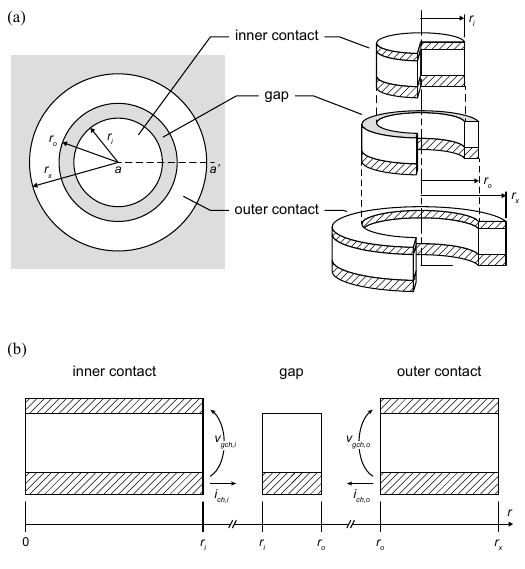}}
	\caption{(a) A circular capacitor and its constituent components.  (b) Cross-sections of the constituent capacitor components along the cutline $a-a'$.}
	\label{fig:circular_cap}
\end{figure}
The radius of the inner contact is $r_i$ and the outer contact's inner and outer radii are $r_o$ and $r_x$, respectively.  For this analysis, it is assumed that the steady-state conductance through the barrier is sufficiently low, but not zero, such that the insulating barrier approximation may be used and, therefore, the channel sheet conductance and barrier admittance are effectively constant with respect to lateral position.  For such a case with circular geometry, \eqref{eq:char_eq_small_sig} becomes
\begin{equation}\label{eq:char_eq_circ}
	\frac{1}{\rnorm} \frac{\partial}{\partial \rnorm} \brk*{ \rnorm \frac{\partial}{\partial \rnorm} v_{gch} \var{\rnorm,t} } - \frac{Y_B \var{\omega} \lambda^2}{\sigmach} v_{gch} \var{\rnorm,t} = 0
\end{equation}
where $\rnorm = r ⁄ \lambda$ is the normalized radial position.  The capacitor has been assumed to be uniform in the azimuthal direction.  By making the substitution
\begin{equation}\label{eq:alpha_def}
	\alpha^2 \! \var{\omega,r} = - \frac{Y_B \var{\omega} \lambda^2}{\sigmach} \rnorm^2 = - \frac{Y_B \var{\omega}}{\sigmach} r^2 = - \mathscr{k}^2 \! \var{\omega} r^2
\end{equation}
with $\mathscr{k} \var{\omega} = \sqrt{ Y_B \var{\omega} / \sigmach }$ being a kind of propagation constant for the small-signal voltage ``wave,'' \eqref{eq:char_eq_circ} can be written in the form of Bessel's differential equation:
\begin{equation}
	\frac{1}{\alpha} \frac{\partial}{\partial \alpha} \brk*{ \alpha \frac{\partial}{\partial \alpha} v_{gch} \var{\alpha,t} } + v_{gch} \var{\alpha,t} = \\
	\frac{\partial^2}{\partial \alpha^2} v_{gch} \var{\alpha,t} + \frac{1}{\alpha} \frac{\partial}{\partial \alpha} v_{gch} \var{\alpha,t} + v_{gch} \var{\alpha,t} = 0 \, ,
\end{equation}
which has solutions of the form
\begin{equation}\label{eq:vgch_circ_general}
	v_{gch} \var{\alpha,t} = A \var{t} J_0 \var{\alpha} + B \var{t} Y_0 \var{\alpha}
\end{equation}
where $J_0 \var{\alpha}$ and $Y_0 \var{\alpha}$ are Bessel functions of the first and second kind, respectively \cite{Zwillinger2003_6_19}.  Given appropriate boundary conditions, e.g., the small-signal gate-channel potential at a specified radius, then the coefficients $A \var{t}$ and $B \var{t}$ can be solved for.  Note, the notation of the frequency dependence of $\mathscr{k} \var{\omega}$ and the frequency and position dependence of $\alpha \var{\omega,r}$ in some of the preceding and subsequent expressions has been dropped simply out of a desire to simplify the notation of the overall expressions.  Additionally, it can be understood that $\alpha$ is effectively a complex, unitless analogue to the radial position (i.e., $\alpha \propto r$) that carries a weighting given by $\mathscr{k}$ (i.e., the barrier admittance and channel sheet conductance).

The entire capacitor structure can be divided for analysis into smaller components: the region under and including inner contact (the inner contact ``capacitor''), the region under and including the outer contact (the outer contact ``capacitor''), and the gap conductor between (Fig.~\ref{fig:circular_cap}a).  For the capacitor associated with the inner contact, the characteristic length is chosen to be the radius $r_i$ with a boundary condition of $v_{gch} \var{r_i,t} = v_{gch,i} \var{t}$, i.e., the small-signal gate-channel voltage across the edge of the inner contact capacitor (Fig.~\ref{fig:circular_cap}b).  At $r = 0$, equivalent to $\alpha = 0$, the Bessel function of the second kind goes to infinity by definition: $\displaystyle\lim_{\alpha \rightarrow \infty} Y_0 \var{\alpha} \rightarrow \infty$.  As the small-signal gate-channel voltage in the capacitor can't be infinite for a finite ``input'' (i.e., boundary condition), the coefficient $B \var{t}$ must be zero.  At the outer edge of the inner contact (i.e., at $r = r_i$ and $\alpha = \alpha_i$), the boundary condition can be applied to \eqref{eq:vgch_circ_general} and
\begin{equation}
	A \var{t} = \frac{1}{J_0 \var{\alpha_i}} v_{gch,i} \var{t} \, .
\end{equation}
The final solution for the small-signal gate-channel potential for the inner contact capacitor is then
\begin{equation}\label{eq:vgch_circ_inner}
	v_{gch} \var{\alpha,t} = \frac{J_0 \var{\alpha}}{J_0 \var{\alpha_i}} v_{gch,i} \var{t} \, .
\end{equation}

In order to determine the admittance or impedance of the inner contact capacitor, the small-signal current flowing through the inner contact capacitor needs to be known.  This current can be evaluated at the gate or the edge of the channel of the inner contact capacitor (Fig.~\ref{fig:circular_cap}b).  Using \eqref{eq:kch_def}, the current in the channel can be evaluated and the small-signal current flowing through the inner contact capacitor is
\begin{equation}\label{eq:ich_circ_inner_gen}
		i_{ch,i} \var{t} = 2 \pi r_i k_{ch} \var{r_i,t} = 2 \pi r_i \eval*{ \frac{\partial}{\partial r} \brk*{ \sigmach v_{gch} \var{r,t} } }_{r = r_i} \\
		= 2 \pi r_i \sigmach \eval*{ \frac{\partial \alpha}{\partial r} }_{r = r_i} \eval*{ \frac{\partial}{\partial \alpha} \brk*{ v_{gch} \var{\alpha,t} } }_{\alpha = \alpha_i} \, .
\end{equation}
Insertion of \eqref{eq:vgch_circ_inner} into \eqref{eq:ich_circ_inner_gen} results in
\begin{equation}\label{eq:ich_circ_inner}
		i_{ch,i} \var{t} = 2 \pi r_i \sigmach \, \imj \mathscr{k} \brk*{ - \frac{J_1 \var{\alpha_i}}{J_0 \var{\alpha_i}} v_{gch,i} \var{t} } \\
		= Y_B \var{\omega} \pi r_i^2 \frac{2}{\alpha_i} \frac{J_1 \var{\alpha_i}}{J_0 \var{\alpha_i}} v_{gch,i} \var{t} \, .
\end{equation}
Equation \eqref{eq:ich_circ_inner}, when divided by the small-signal gate-channel voltage ``applied'' across the inner contact capacitor (i.e., the boundary condition $v_{gch,i} \var{t}$), directly leads to an admittance of
\begin{equation}\label{eq:Y_inner}
	Y_i \var{\omega} = \frac{i_{ch,i} \var{t}}{v_{gch,i} \var{t}} = Y_B \var{\omega} \pi r_i^2 \frac{2}{\alpha_i} \frac{J_1 \var{\alpha_i}}{J_0 \var{\alpha_i}} \, .
\end{equation}
The associated impedance is simply the inverse of the admittance: $Z_i \var{\omega} = 1 ⁄ Y_i \var{\omega}$.

For the capacitor associated with the outer contact, the characteristic length is chosen to be the inner radius $r_o$ with a boundary condition of $v_{gch} \var{r_o,t} = v_{gch,o} \var{t}$, i.e., the small-signal gate-channel voltage across the inner edge of the outer contact capacitor (Fig.~\ref{fig:circular_cap}b).  At the outer edge of the outer contact, $r = r_x$ (equivalent $\alpha = \alpha_x = \sqrt{- Y_B \var{\omega} / \sigmach} r_x = \imj \mathscr{k} r_x$), the total lateral sheet current density in the channel (\eqref{eq:kch_def} inserting \eqref{eq:vgch_circ_general}) necessarily goes to zero:
\begin{equation}
		k_{ch} \var{r_x,t} = \eval*{ \frac{\partial}{\partial r} \brk*{ \sigmach v_{gch} \var{r,t} } }_{r = r_x} \\
		= - \imj \sigmach \mathscr{k} \brk^1( A \var{t} J_1 \var{\alpha_x} \\
		+ B \var{t} Y_1 \var{\alpha_x} \brk^1) = 0 \, ,
\end{equation}
therefore
\begin{equation}
	B \var{t} = - \frac{J_1 \var{\alpha_x}}{Y_1 \var{\alpha_x}} A \var{t} \, .
\end{equation}
At the inner radius of the outer contact ($r = r_o$, $\alpha = \alpha_o$), the boundary condition can be applied and
\begin{equation}
	A \var{t} = \frac{1}{J_0 \var{\alpha_o} - Y_0 \var{\alpha_o} J_1 \var{\alpha_x} / Y_1 \var{\alpha_x}} v_{gch,o} \var{t} \, .
\end{equation}
The final solution for the small-signal gate-channel potential for the outer contact capacitor is then
\begin{equation}\label{eq:vgch_circ_outer}
	v_{gch} \var{\alpha,t} = \frac{J_0 \var{\alpha} - Y_0 \var{\alpha} J_1 \var{\alpha_x} / Y_1 \var{\alpha_x}}{J_0 \var{\alpha_o} - Y_0 \var{\alpha_o} J_1 \var{\alpha_x} / Y_1 \var{\alpha_x}} v_{gch,o} \var{t} \, .
\end{equation}

The admittance (impedance) of the outer contact capacitor is found in the same manner as the inner contact capacitor, except here the current can be evaluated at the gate or the inner edge of the channel of the outer contact capacitor.  Like \eqref{eq:ich_circ_inner_gen}, the current in the channel can be evaluated and the small-signal current flowing through the outer contact capacitor is
\begin{equation}\label{eq:ich_circ_outer}
	i_{ch,o} \var{t} = - 2 \pi r_o k_{ch} \var{r_o,t} \\
	= 2 \pi \sigmach \alpha_o \frac{J_1 \var{\alpha_o} - Y_1 \var{\alpha_o} J_1 \var{\alpha_x} / Y_1 \var{\alpha_x}}{J_0 \var{\alpha_o} - Y_0 \var{\alpha_o} J_1 \var{\alpha_x} / Y_1 \var{\alpha_x}} v_{gch,o} \var{t}
\end{equation}
after insertion of \eqref{eq:vgch_circ_outer}.  Note, the negative sign in \eqref{eq:ich_circ_outer} is a result of $i_{ch,o} \var{t}$ being defined to flow in the opposite direction as $k_{ch} \var{r,t}$ at $r = r_o$, $\alpha = \alpha_o$ (Fig.~\ref{fig:circular_cap}b).  The admittance of the outer contact capacitor is then
\begin{equation}\label{eq:Y_outer}
		Y_o \var{\omega} = \frac{i_{ch,o} \var{t}}{v_{gch,o} \var{t}} \\
		= 2 \pi \sigmach \alpha_o \frac{J_1 \var{\alpha_o} - Y_1 \var{\alpha_o} J_1 \var{\alpha_x} / Y_1 \var{\alpha_x}}{J_0 \var{\alpha_o} - Y_0 \var{\alpha_o} J_1 \var{\alpha_x} / Y_1 \var{\alpha_x}} \, .
\end{equation}
The associated impedance is simply the inverse of the admittance: $Z_o \var{\omega} = 1 ⁄ Y_o \var{\omega}$.

The admittance presented in \eqref{eq:Y_outer} is that for a ``finite'' outer contact.  In the case of an ``infinite'' outer contact, where the outer radius of the outer contact becomes very, very large (i.e., $r_x \rightarrow \infty$, $\abs{ \alpha_x } \rightarrow \infty$), the Bessel functions can be approximated by
\begin{equation}\label{eq:Bessel_hi_alpha_approx}
	J_n \var{\alpha} \approx \sqrt{ \frac{2}{\pi \alpha} } \cos{\brk*{\alpha - \frac{n\pi}{2} - \frac{\pi}{4}}} \quad \text{and} \\
	\quad
	Y_n \var{\alpha} \approx \sqrt{ \frac{2}{\pi \alpha} } \sin{\brk*{\alpha - \frac{n\pi}{2} - \frac{\pi}{4}}}
\end{equation}
for $\abs{\arg{\brk{\alpha}}} < \pi$  \cite{Zwillinger2003_6_19}, which is always true for an admittance composed of any combination of conductances and capacitances.  Therefore, in the limit that $\abs{\alpha_x} \rightarrow \infty$, the ratio $J_1 \var{\alpha_x} ⁄ Y_1 \var{\alpha_x}$ reduces to
\begin{equation}
		\lim_{\abs{\alpha_x} \rightarrow \infty} \frac{J_1 \var{\alpha_s}}{Y_1 \var{\alpha_x}} \approx \lim_{\abs{\alpha_x} \rightarrow \infty} \frac{ \cos{\brk*{\alpha_x - \frac{3}{4} \pi}} }{ \sin{\brk*{\alpha_x - \frac{3}{4} \pi}} } \\
		\approx \lim_{\abs{\alpha_x} \rightarrow \infty} \imj \frac{ e^{\imj \brk*{\alpha_x - \frac{3}{4} \pi}} + e^{- \imj \brk*{\alpha_x - \frac{3}{4} \pi}} }{ e^{\imj \brk*{\alpha_x - \frac{3}{4} \pi}} - e^{- \imj \brk*{\alpha_x - \frac{3}{4} \pi}} } \\
		\approx \lim_{\abs{\alpha_x} \rightarrow \infty} \imj \frac{ e^{\imj \brk*{\alpha_x - \frac{3}{4} \pi}} }{ e^{\imj \brk*{\alpha_x - \frac{3}{4} \pi}} } \approx \imj \, .
\end{equation}
And the admittance of an infinite outer contact capacitor becomes
\begin{equation}\label{eq:Y_outer_inf}
		Y_o^{\brk{\infty}} \var{\omega} = 2 \pi \sigmach \alpha_o \frac{ J_1 \var{\alpha_o} - \imj Y_1 \var{\alpha_o} }{ J_0 \var{\alpha_o} - \imj Y_0 \var{\alpha_o} } \\
		= 2 \pi \sigmach \alpha_o \frac{ H_1^{\brk{2}} \var{\alpha_o} }{ H_0^{\brk{2}} \var{\alpha_o} }
\end{equation}
where $H_n^{\brk{2}} \var{\alpha} = J_n \var{\alpha} - \imj Y_n \var{\alpha}$ is the Hankel function of the second kind \cite{Zwillinger2003_6_19}.  The associated impedance is again simply $Z_o^{\brk{\infty}} \var{\omega} =1 ⁄ Y_o^{\brk{\infty}} \var{\omega}$.

For capacitors where the outer contact is in direct electrical contact with the channel (e.g., the outer contact is ohmic), the impedance of the outer contact reduces to the contact resistance of that contact:
\begin{equation}\label{eq:Z_outer_ohm}
	Z_o \var{\omega} = \frac{R_C}{2 \pi r_o} \approx \frac{1}{2 \pi r_o} \frac{L_T}{\sigmach}
\end{equation}
where $R_C \approx L_T ⁄ \sigmach$  is the contact resistance ($\Omega\cdot$mm) and $L_T$ is the transfer length of the contact.  The approximation in the above expression is valid for outer contacts with ``lengths'' (i.e., $r_x - r_o$) much greater than the transfer length: $r_x - r_o \gg L_T$ \cite{Willis1987}.

The impedance of the gap between the inner and outer contacts is simply the resistance of the annular section of the channel between the contacts:
\begin{equation}\label{eq:Z_gap}
	Z_{gap} \var{\omega} = \frac{1}{2 \pi \sigmach} \ln{\brk*{\frac{r_o}{r_i}}} \, .
\end{equation}
The total impedance of the full capacitor structure is the sum of the constituent components:
\begin{equation}\label{eq:ZT_def}
	Z_T \var{\omega} = Z_i \var{\omega} + Z_{gap} \var{\omega} + Z_o \var{\omega} \, .
\end{equation}
Equation \eqref{eq:ZT_def} provides the total impedance of the capacitor.  However, in practical measurements of such structures and devices, there is often an additional series resistance ($R_S$) due to probe resistance, at a minimum, and possibly additional instrument and cable resistances.  This ``extrinsic'' series resistance can simply be added to the total impedance of the capacitor in the analysis, e.g., $Z_T \var{\omega} + R_S$.

The total impedance can be expressed in terms of an equivalent resistance and capacitance connected in series:
\begin{equation}
	Z_T \var{\omega} \equiv R_T \var{\omega} - \imj \frac{1}{\omega C_T \var{\omega}}
\end{equation}
with
\begin{subequations}\label{eq:RT_CT}
	\begin{equation}\label{eq:RT}
		R_T \var{\omega} = \Real{\brk[s]*{ Z_T \var{\omega} }}
	\end{equation}
	\begin{equation}\label{eq:CT}
		C_T \var{\omega} = - \frac{1}{ \omega \Imag{\brk[s]*{ Z_T \var{\omega} }} } \, .
	\end{equation}
\end{subequations}
This representation (i.e., a series equivalent circuit), as opposed to a parallel equivalent circuit, is a better representative model of the overall behavior of the capacitor for capacitors with a nearly insulating barrier.

\begin{figure}[!htb]
	\centerline{\includegraphics[width=15.0cm]{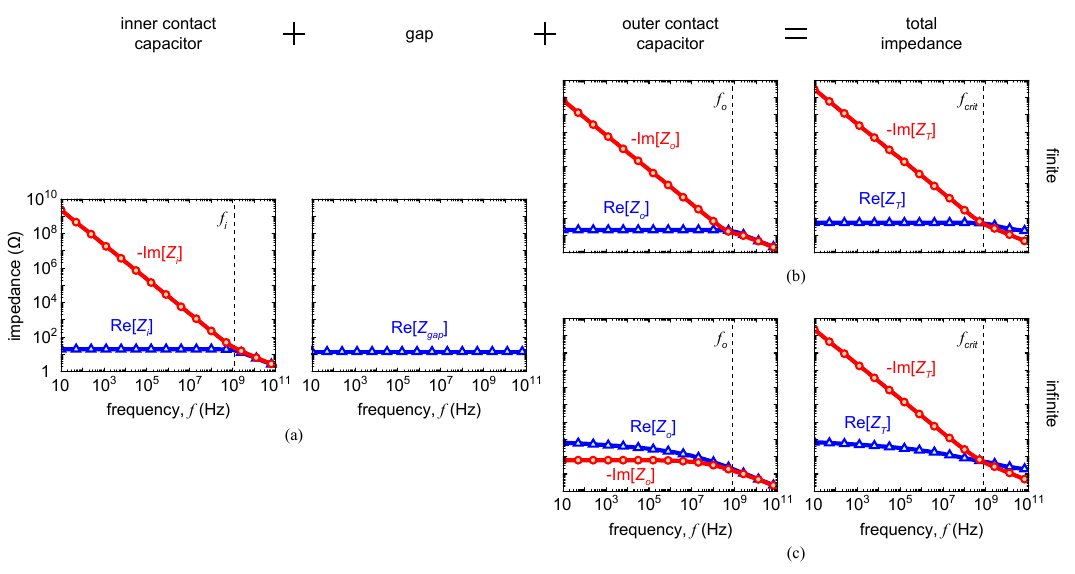}}
	\caption{Plots of the real and imaginary components of the constituent component impedances and total impedance of a circular capacitor as shown in Fig. 2.  (a) The impedance of the inner contact capacitor ($Z_i$, left) and gap ($Z_{gap}$, right), which are independent of the outer contact size.  The impedance of the outer contact capacitor ($Z_o$, left) and the sum total ($Z_i + Z_{gap} + Z_o = Z_T$, right) for (b) a finite outer contact and (c) an infinite outer contact.}
	\label{fig:circular_cap_components}
\end{figure}

For reference, the component and total impedances of a capacitor with a finite outer contact and an infinite outer contact with a perfectly insulating barrier ($G_B \var{\omega,\vec{r}} = 0$) is plotted in Fig.~\ref{fig:circular_cap_components}.
The relevant capacitor properties are listed in Table~\ref{tab:cap_parameters}.  As can be seen from the figure, the key and only difference between the full finite and infinite outer contact capacitors is the behavior of the capacitor associated with the outer contact, as should be expected given that the other components (i.e., inner contact capacitor and gap) remain wholly unchanged.
\begin{table}[!htb]
	\centering
	\caption{Perfectly Insulating Barrier Capacitor Properties}
	\label{tab:cap_parameters}
	\small
\begin{tabular}{| l | c |}
	\hline
	inner contact radius, $r_i$ & $50$~{\textmu}m \\
	outer contact inner radius, $r_o$ & $1.2 r_i = 60$~{\textmu}m \\
	outer contact outer radius, $r_x$ & $2.2 r_i = 110$~{\textmu}m \\
	barrier relative permittivity, $\epsilon_r$ & $10$ \\
	barrier thickness, $d_B$ & $100$~nm \\
	barrier capacitance, $C_B = \epsilon_r \epsilon_0 / d_B$ & $88.53$~nF/cm$^2$ \\
	steady-state channel sheet conductance, $\sigmach$ & $500$~S \\
	\hline
\end{tabular}
\end{table}

\section{Model Behavior}
\subsection{Frequency Response}
As can be observed in Fig.~\ref{fig:circular_cap_components}, there exists a clear frequency dependent behavior that can be split into two distinct regions: a low frequency region and a high frequency region, separated by some critical frequency, $f_{crit}$.  To understand this behavior, it is useful to examine its low and high frequency limits.

The Bessel functions of the first and second kind seen in the admittances/impedances of the circular capacitor can be expressed in terms of series:
\begin{subequations}
	\begin{equation}
		J_n \var{\alpha} = \sum_{m=0}^{\infty} \frac{\brk{-1}^m}{m! \, \Gamma \var{m+n+1}} \brk*{ \frac{\alpha}{2} }^{2m+n}
	\end{equation}
	\begin{equation}
		Y_n \var{\alpha} = \frac{2}{\pi} \ln{\brk*{\frac{\alpha}{2}}} J_n \var{\alpha} \\
		- \frac{1}{\pi} \sum_{m=0}^{\infty} \brk[s]*{ \psi \var{m+1} + \psi \var{m+n+1} } \frac{\brk{-1}^m}{m! \brk{m+n}!} \brk*{\frac{\alpha}{2}}^{2m+n} \\
		- \frac{1}{\pi} \sum_{m=0}^{n-1} \frac{\brk{n-m-1}!}{m!} \brk*{\frac{\alpha}{2}}^{2m-n}
	\end{equation}
\end{subequations}
where $\psi \var{\zeta}$ is the digamma function \cite{Zwillinger2003_6_19}.  For small values of $\alpha$ (e.g., at low frequencies), the Bessel functions may be approximated by only a few terms of the series.  Therefore, it can be shown that the approximate forms of the Bessel functions for $n = 0$ are
\begin{subequations}
	\begin{equation}
		J_0 \var{\alpha} \approx 1 - \brk*{ \frac{\alpha}{2} }^2 = 1 - \frac{\alpha^2}{4}
	\end{equation}
	\begin{equation}
		Y_0 \var{\alpha} \approx \frac{2}{\pi} \brk[s]*{ \ln{\brk*{ \frac{\alpha}{2} }} + \gamma } J_0 \var{\alpha}
	\end{equation}
\end{subequations}
where $\gamma = 0.57221 \cdots$ is Euler’s constant, and for $n = 1$ are
\begin{subequations}
	\begin{equation}
		J_1 \var{\alpha} \approx \frac{\alpha}{2} - \frac{1}{2} \brk*{ \frac{\alpha}{2} }^3 = \frac{\alpha}{2} \brk*{ 1 - \frac{\alpha^2}{8} }
	\end{equation}
	\begin{equation}
		Y_1 \var{\alpha} \approx \frac{2}{\pi} \brk[c]*{ \brk[s]*{ \ln{\brk*{\frac{\alpha}{2}}} + \gamma } J_1 \var{\alpha} - \frac{1}{\alpha} J_0 \var{\alpha} - \frac{\alpha}{2} }
	\end{equation}
\end{subequations}
for $\abs{\alpha^2} \ll 1$, which is equivalent to $\abs{ Y_B \var{\omega} }r^2 \ll \sigmach$, reasonably valid for an approximately insulating barrier at low frequency.  The low frequency approximations of the various components of the overall capacitor (e.g., the inner contact capacitor, the outer contact capacitor) are found by applying these approximations for the Bessel functions to the respective impedances.

For the inner contact capacitor:
\begin{equation}\label{eq:Zi_lo_initial}
		Z_i^{\brk{lo}} \var{\omega} \approx \frac{1}{Y_B \var{\omega} \pi r_i^2} \frac{1 - \frac{\alpha_i^2}{4}}{1 - \frac{\alpha_i^2}{8}} \\
		\approx \frac{1}{Y_B \var{\omega} \pi r_i^2} \frac{ 1 - \frac{\alpha_i^2}{4} - \frac{\brk{\alpha_i^*}^2}{8} + \frac{\abs{\alpha_i^2}^2}{32} }{ \abs*{1 - \frac{\alpha_i^2}{8}}^2 } \\
		\approx \frac{1}{Y_B \var{\omega} \pi r_i^2} \brk[s]*{ 1 - \frac{\alpha_i^2}{4} - \frac{\brk{\alpha_i^*}^2}{8} }
\end{equation}
where $\alpha_i^2 = - Y_B \var{\omega} r_i^2 / \sigmach$ is by definition a complex quantity and $\brk{\alpha_i^*}^2 = \brk{\alpha_i^2}^* = - Y_B^* \var{\omega} r_i^2 / \sigmach$ is its complex conjugate.  At first, \eqref{eq:Zi_lo_initial} looks overly complicated.  However, when the barrier conductance is sufficiently low relative to the barrier susceptance (i.e., $G_B \var{\omega} \ll \omega C_B \var{\omega}$), congruent with an approximately insulating barrier even at low frequencies, then $Y_B^* \var{\omega} \approx - Y_B \var{\omega}$ and consequently $\brk{\alpha_i^*}^2 \approx - \alpha_i^2$; \eqref{eq:Zi_lo_initial} simplifies to
\begin{equation}\label{eq:Zi_lo}
		Z_i^{\brk{lo}} \var{\omega} \approx \frac{1}{Y_B \var{\omega} \pi r_i^2} \brk*{ 1 - \frac{\alpha_i^2}{8} } \\
		\approx \frac{1}{Y_B \var{\omega} \pi r_i^2} + \frac{1}{8 \pi \sigmach} \, .
\end{equation}
What is striking about \eqref{eq:Zi_lo} is that the low frequency impedance of the inner contact capacitor is essentially and simply what one would expect from a low frequency approximate model: the series connection of the lumped element representations of the impedance of the inner contact barrier ($1 / Y_B \var{\omega} \pi r_i^2$) and the resistance of a conductive disc associated with the channel beneath the inner contact ($1 / 8 \pi \sigmach$), where current flows between one face of the disc and its edge \cite{Phillips1962}.

Using the same procedure for the outer contact capacitor, its low frequency approximation is
\begin{equation}\label{eq:Zo_lo}
	Z_o^{\brk{lo}} \var{\omega} \approx \frac{1}{Y_B \var{\omega} \pi \brk*{r_x^2 - r_o^2}} \\
	+ \frac{1}{8 \pi \sigmach} \brk[s]*{ \frac{4 r_x^4}{\brk*{r_x^2 - r_o^2}^2} \ln{\brk*{\frac{r_x}{r_o}}} - \frac{2 r_x^2}{r_x^2 - r_o^2} - 1 } \, ,
\end{equation}
which, like \eqref{eq:Zi_lo}, is simply the series connection of the lumped element representations of the impedance of the outer contact barrier and the resistance of a conductive annular ring associated with the channel beneath the outer contact, where current flows between the inner edge of the ring and one face.  The impedance of an infinite outer contact (i.e., $r_x \rightarrow \infty$) capacitor is
\begin{equation}\label{eq:Zo_lo_inf}
	Z_o^{\brk{\infty, lo}} \var{\omega} \approx \frac{1}{2 \pi \sigmach} \brk[s]*{ \frac{1}{2} \ln{\brk*{ \frac{1}{2} \frac{8 \pi \sigmach}{Y_B \var{\omega} \pi r_o^2} }} - \gamma } \, .
\end{equation}

The high frequency approximations of the various components of the overall capacitor (e.g., the inner contact capacitor, the outer contact capacitor) can be found by applying the approximations from \eqref{eq:Bessel_hi_alpha_approx} to the expressions for impedance.  For the inner contact capacitor:
\begin{equation}
	Z_i^{\brk{hi}} \var{\omega} \approx \frac{\sqrt{2}}{Y_B \var{\omega} \pi r_i^2} \sqrt{ \frac{Y_B \var{\omega} \pi r_i^2}{8 \pi \sigmach} } \, .
\end{equation}
For the outer contact capacitor, irrespective of its outer radius ($r_x$):
\begin{equation}\label{eq:Zo_hi}
	Z_o^{\brk{hi}} \var{\omega} \approx \frac{\sqrt{2}}{8 \pi \sigmach} \sqrt{ \frac{8 \pi \sigmach}{Y_B \var{\omega} \pi r_o^2} } \, .
\end{equation}
Equation \eqref{eq:Zo_hi} is valid for both finite and infinite outer contact capacitors because even an outer contact with a finite outer radius appears infinite at very high frequencies and associated short wavelengths.

In the expressions for the low frequency and high frequency limits of the various impedances, two terms repeatedly appear.  The resistance of a disc associated with the channel, $1 / 8 \pi \sigmach$, and the admittance of the barrier above said disc, $Y_B \var{\omega} \pi r_m^2$ (with $m = i$ for the inner contact radius and $m = o$ for the inner radius of the outer contact).  Furthermore, this resistance and admittance (impedance) pair, or similar modified by appropriate geometric factors, appear in series within the low frequency approximations (see \eqref{eq:Zi_lo} and \eqref{eq:Zo_lo}), suggesting a series $RC$-like equivalent circuit structure.  Taking the product of this admittance and resistance, and additionally assuming again the barrier susceptance dominates the admittance (i.e., $\omega C_B \var{\omega} \gg G_B \var{\omega}$), directly results in an expression for an $RC$-like time constant, $\tau_m \var{\omega}$:
\begin{equation}\label{eq:tau_m}
	\frac{Y_B \var{\omega} \pi r_m^2}{8 \pi \sigmach} \approx \frac{\imj \omega C_B \var{\omega} \pi r_m^2}{8 \pi \sigmach} = \imj \omega \tau_m \var{\omega}
\end{equation}
where $\tau_m \var{\omega} = C_B \var{\omega} \pi r_m^2 / 8 \pi \sigmach$ is the $RC$-like time constant associated with the inner ($m = i$) and outer ($m = o$) contact capacitors.  Strictly, due to the possible frequency dependence of the barrier capacitance, $C_B \var{\omega}$, this time constant is similarly frequency dependent.  However, in many cases, the barrier capacitance will be effectively constant over a wide range of frequencies, resulting in an approximately constant value for the time constant.

The frequencies given by $f_m = 1 ⁄ 2 \pi \tau_m \var{\omega}$ (i.e., $\omega \tau_m \var{\omega} = 1$) lie at the border between the low-frequency and high-frequency approximations for their respective capacitors and effectively define the maximum and minimum limits in frequency of those approximations, respectively.  For the full capacitor, this critical frequency is dictated by the lower of the two frequencies:
\begin{equation}\label{eq:fcrit}
	f_{crit} \equiv f_o = \frac{1}{2 \pi \tau_o \var{\omega}} = \frac{1}{2 \pi} \frac{8 \pi \sigmach}{C_B \var{\omega} \pi r_o^2}
\end{equation}
as $r_o > r_i$ and therefore $1 / 2 \pi \tau_o \var{\omega} = f_o < f_i = 1 / 2 \pi \tau_i \var{\omega}$.

\begin{figure}[!htb]
	\centerline{\includegraphics[width=15.0cm]{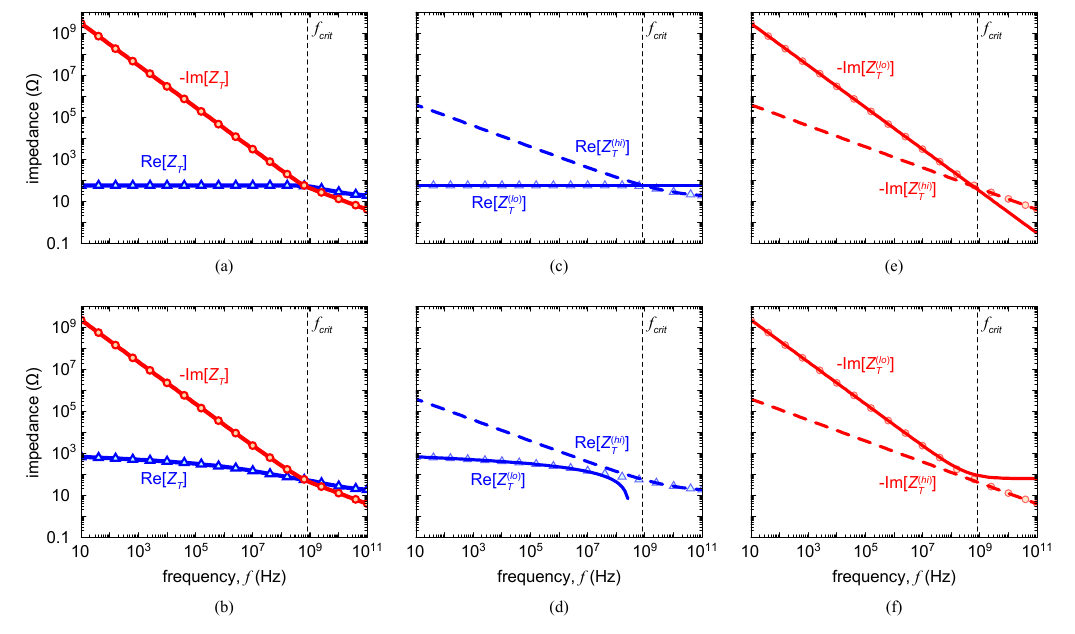}}
	\caption{The real and imaginary components of the total impedance for a circular capacitor with (a) a finite outer contact and (b) an infinite outer contact.  The low frequency (solid lines) and high frequency (dashed lines) approximations are shown superimposed on the real component (triangles, c and d) and the imaginary component (circles, e and f) of the finite outer contact capacitor and the infinite outer contact capacitor, respectively.  The critical frequency, $f_{crit}$, is indicated, demarking the border between the applicable ranges for the low and high frequency approximations.}
	\label{fig:circular_cap_approximations}
\end{figure}

Fig.~\ref{fig:circular_cap_approximations} shows again the total impedances of a capacitor with a finite outer contact and an infinite outer contact with a perfectly insulating barrier ($G_B \var{\omega, \vec{r}}$), now with their respective low frequency and high frequency approximations.
In the figure, it can be seen that the expressions in \eqref{eq:Zi_lo} - \eqref{eq:Zo_hi} are excellent approximations to the exact expression for the impedance of the circular capacitor, with $f_{crit}$ of \eqref{eq:fcrit}, marking the effective border between the applicable domains of these approximations.  To be clear, the exact expressions for the various components of the full capacitor (e.g., \eqref{eq:Y_inner}, \eqref{eq:Y_outer}, \eqref{eq:Y_outer_inf}, \eqref{eq:Z_outer_ohm}, and \eqref{eq:Z_gap}) remain valid for all frequencies.  The different time constants, \eqref{eq:tau_m}, and frequencies, \eqref{eq:fcrit}, simply highlight where the behavior of the capacitor distinctly changes and the application limits of the low and high-frequency approximations of \eqref{eq:Zi_lo} - \eqref{eq:Zo_hi}.  Furthermore, the approximations themselves primarily provide an intuitive representation for visualizing the various impedances, but may also be useful in rudimentary parameter extraction.

\subsection{Dielectric Response}
As noted previously, capacitors are often used to evaluate the electronic and/or dielectric properties of semiconducting or dielectric materials.  To understand the behavior of practical capacitors and how that impacts the evaluation of the constituent materials, it is useful to examine an uncomplicated capacitor consisting of a single dielectric layer as the barrier material.  This simplifies the expression of the barrier admittance, which in turn simplifies the analysis of the capacitor, however the conclusions reached here are equally applicable to more complex structures.  Furthermore, in an effort to examine the larger, overall behavior of the capacitor, it is useful to express the admittance of the barrier solely as a capacitance with a complex relative permittivity:
\begin{equation}\label{eq:Y_B_complex_perm}
	Y_B \var{\omega} = \imj \omega C_B \var{\omega} = \imj \omega \frac{\epsilon_r \var{\omega} \epsilon_0}{d_B}
\end{equation}
where $\epsilon_r \var{\omega}$ is the complex relative permittivity:
\begin{equation}\label{eq:complex_perm}
	\epsilon_r \var{\omega} = \epsilon_r^{\prime} \var{\omega} - \imj \epsilon_r^{\prime \prime} \var{\omega}
\end{equation}
with $\epsilon_r^{\prime} \var{\omega}$ and $\epsilon_r^{\prime \prime} \var{\omega}$ being the real and imaginary components of the full, complex relative permittivity, respectively; $\epsilon_0 = 8.854 \times 10^{-12}$~F/m being the vacuum permittivity; and $d_B$ being the thickness of the barrier.  There is no effective difference between how the admittance is presented in \eqref{eq:Y_B_complex_perm} and \eqref{eq:YB_def}, as any conductance in the barrier can be expressed equivalently through the imaginary component of the relative permittivity, e.g., $\epsilon_r^{\prime \prime} = \sigma_B / \omega \epsilon_0$ with $\sigma_B$ being the conductivity (S/cm) of the barrier.

The dissipation factor or loss tangent characterizes the inherent power dissipation or loss in a given electrical element, as the name suggests, and is given by the ratio of the power dissipated to the power stored in said element.  For a generic impedance (admittance) of $Z = R + \imj X$ ($Y = Z^{-1} = G + \imj B$), the associated loss tangent, $\tan{\var{\delta}}$, of that impedance (admittance) is
\begin{equation}\label{eq:tan_d}
	\tan{\var{\delta}} = - \frac{R}{X} = \frac{G}{B} \, .
\end{equation}
Equation \eqref{eq:tan_d} can equally be applied to full electronic devices, such as a capacitor, or to a single material layer.  When applied to a single dielectric layer as described by \eqref{eq:Y_B_complex_perm} and \eqref{eq:complex_perm}, \eqref{eq:tan_d} gives the loss tangent of the material, notated as $\tan{\var{\delta_\epsilon}}$ to distinguish it from the more generic expression in \eqref{eq:tan_d}:
\begin{equation}\label{eq:tan_d_matl}
	\tan{\var{\delta_\epsilon}} = \frac{\epsilon_r^{\prime \prime} \var{\omega}}{\epsilon_r^{\prime} \var{\omega}} \, .
\end{equation}
As will be shown, the loss tangent of a full capacitor can be markedly different from that of the constituent barrier material; and erroneously attributing \eqref{eq:tan_d_matl} to the full capacitor behavior can result in incorrect interpretations with regard to the dielectric properties of the constituent materials.

\begin{figure}[!htb]
	\centerline{\includegraphics[width=15.0cm]{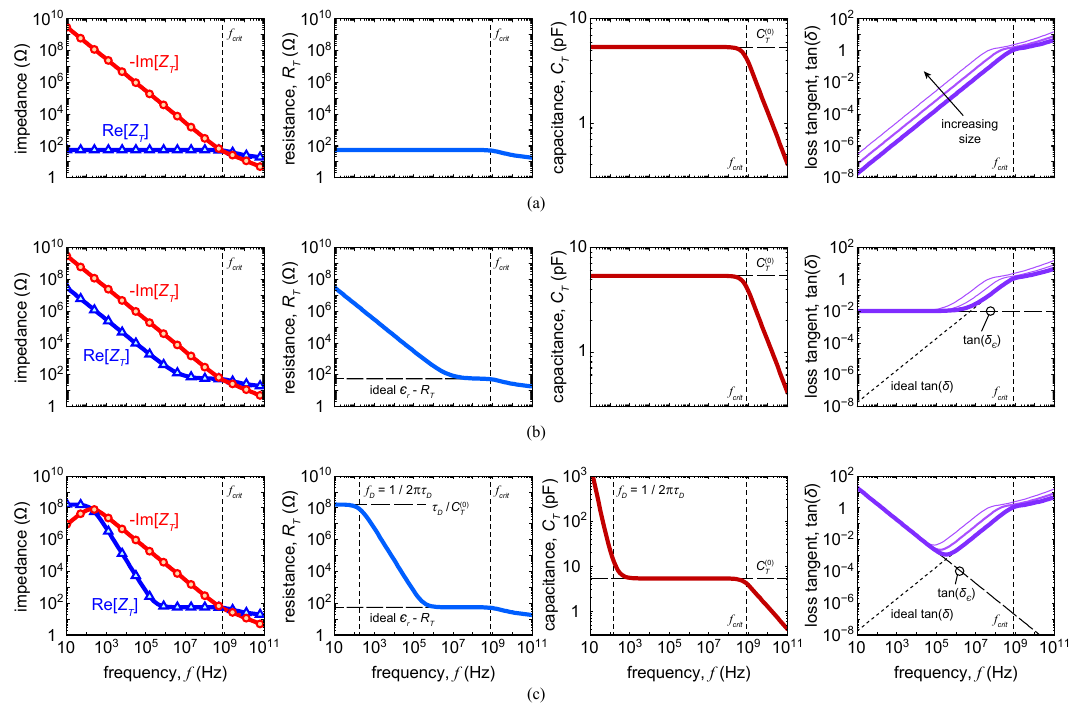}}
	\caption{Plots of impedance, equivalent series resistance ($R_T \var{\omega}$, \eqref{eq:RT}), equivalent series capacitance ($C_T \var{\omega}$, \eqref{eq:CT}), and loss tangent of \eqref{eq:tan_d} for a circular capacitor with (a) a constant and real relative permittivity ($\epsilon_r \var{\omega} = \epsilon_r^{\prime}$), (b) a constant but complex relative permittivity ($\epsilon_r \var{\omega} = \epsilon_r^{\prime} - \imj \epsilon_r^{\prime \prime}$), and (c) a constant, real relative permittivity with a constant barrier conductivity ($\epsilon_r \var{\omega} = \epsilon_r^{\prime} - \imj \sigma_B / \omega \epsilon_0$).  In the plots of equivalent series capacitance ($C_T \var{\omega}$), the approximate low-frequency equivalent series capacitance (i.e., $C_T^{\brk{0}} \var{\omega}$ of \eqref{eq:CT_0}) is plotted for comparison.  In the plots of loss tangent ($\tan{\var{\delta}}$), the loss tangent of the material ($\tan{\var{\delta_\epsilon}}$) is plotted for comparison.}
	\label{fig:circular_cap_dielectric}
\end{figure}

Fig.~\ref{fig:circular_cap_dielectric} shows the total impedance, equivalent series resistance and capacitance of \eqref{eq:RT_CT}, and loss tangent of a circular capacitor with finite outer contact for three exemplar cases: (a) a constant and real relative permittivity (the ``ideal'' dielectric case, $\epsilon_r \var{\omega} = \epsilon_r^{\prime}$), (b) a constant but complex relative permittivity ($\epsilon_r \var{\omega} = \epsilon_r^{\prime} - \imj \epsilon_r^{\prime \prime}$), and (c) a constant, real relative permittivity with a constant barrier conductivity, $\sigma_B$ ($\epsilon_r \var{\omega} = \epsilon_r^{\prime} - \imj \sigma_B / \omega \epsilon_0$).  The capacitor properties are the same as in Table~\ref{tab:cap_parameters} with the real relative permittivity being $\epsilon_r^{\prime} = 10$, the additional imaginary relative permittivity for the second case being $\epsilon_r^{\prime \prime} = 0.01 \epsilon_r^{\prime} = 0.1$, and the barrier conductivity for the third case being $\sigma_B = 10^{-9}$~S/cm.
In the plots for loss tangent, $\tan{\var{\delta}}$, three capacitor sizes are shown: inner contact radii of $r_i = 50$~{\textmu}m (the value in Table~\ref{tab:cap_parameters}), $r_i = 100$~{\textmu}m, and $r_i = 200$~{\textmu}m all with outer contact inner and outer radii with the same proportionalities as shown in Table~\ref{tab:cap_parameters} (i.e., $r_o = 1.2 r_i$ and $r_x = 2.2 r_i$).

\begin{figure}[!htb]
	\centerline{\includegraphics[width=15.0cm]{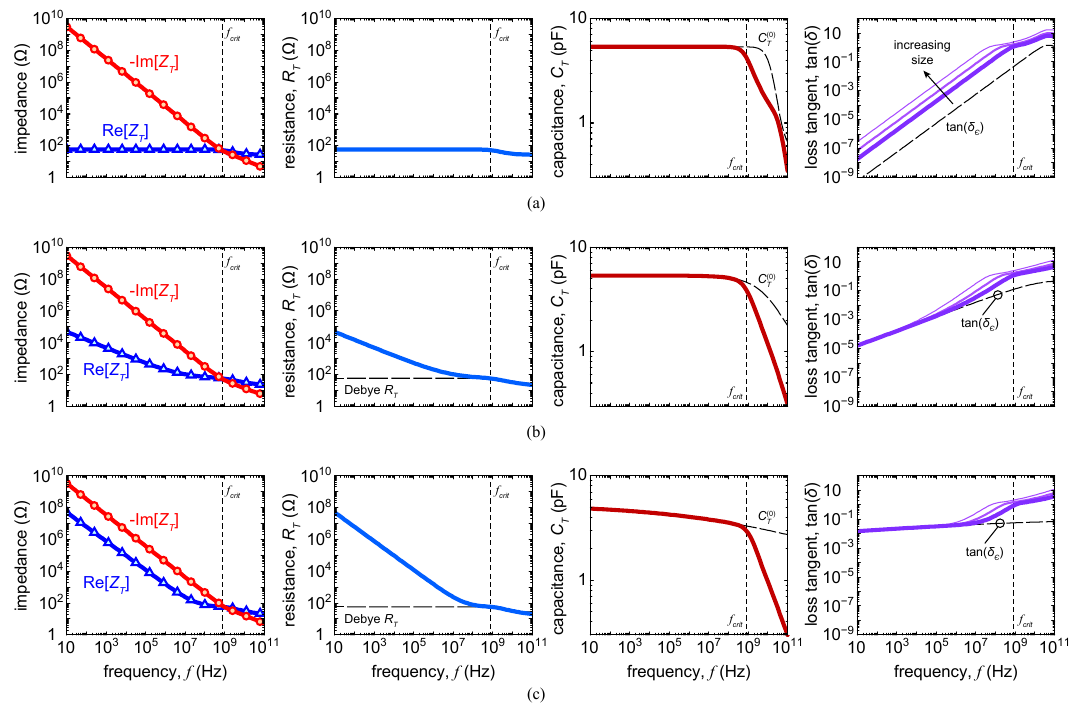}}
	\caption{Plots of impedance, equivalent series resistance, equivalent series capacitance, and loss tangent for a circular capacitor with the Cole-Cole dielectric response of eq. 63 and with $\epsilon_{r,s} = 10$, $\epsilon_{r,\infty} = 1$, $\tau = 10^{-11}$~sec, and $\alpha_{cc}$ equal to (a) 1, (b) 0.5, and (c) 0.1.  In the plots of equivalent series capacitance ($C_T \var{\omega}$), the approximate low-frequency equivalent series capacitance (i.e., $C_T^{\brk{0}} \var{\omega}$ of \eqref{eq:CT_0}) is plotted for comparison.  In the plots of loss tangent ($\tan{\var{\delta}}$), the loss tangent of the material ($\tan{\var{\delta_\epsilon}}$) is plotted for comparison.}
	\label{fig:circular_cap_Cole_Cole}
\end{figure}

Fig.~\ref{fig:circular_cap_Cole_Cole} shows a fourth exemplar: a circular capacitor with finite outer contact with a Cole-Cole dielectric response \cite{ColeCole1941}:
\begin{equation}\label{eq:Cole_Cole}
	\epsilon_r \var{\omega} = \epsilon_{r,\infty} + \frac{\epsilon_{r,s} - \epsilon_{r,\infty}}{1 + \brk*{\imj \omega \tau}^{\alpha_{cc}}}
\end{equation}
where $\epsilon_{r,s}$ and $\epsilon_{r,\infty}$ are the static and high-frequency relative permittivities of the dielectric, respectively, $\tau$ is a characteristic time constant associated with the transition between the static and high-frequency relative permittivities, and $\alpha_{cc}$ is an empirical shape parameter with a value between 0 and 1 ($\alpha_{cc}=0$ corresponds to a constant, real relative permittivity with a value equal to the static value and $\alpha_{cc}=1$ corresponds to the classic Debye dielectric response).  In the plots of Fig.~\ref{fig:circular_cap_Cole_Cole}, $\epsilon_{r,s} = 10$, $\epsilon_{r,\infty} = 1$, $\tau = 10^{-11}$~sec, and $\alpha_{cc}$ is set to 1, 0.5, and 0.1.
As in Fig.~\ref{fig:circular_cap_dielectric}, in the plots for loss tangent, $\tan{\var{\delta}}$, three capacitor sizes are shown: inner contact radii of $r_i = 50$~{\textmu}m, $r_i = 100$~{\textmu}m, and $r_i = 200$~{\textmu}m all with outer contact inner and outer radii with the same proportionalities as shown in Table~\ref{tab:cap_parameters} (i.e., $r_o = 1.2 r_i$ and $r_x = 2.2 r_i$).

A number of observations can be made from Figs.~\ref{fig:circular_cap_dielectric} and \ref{fig:circular_cap_Cole_Cole}, which can provide insight into the electronic and dielectric properties of the capacitor.  Much of the observable behavior that is useful falls below the critical frequency of \eqref{eq:fcrit} in the ``low frequency'' range.  Above the critical frequency, in the ``high frequency'' range, the real and imaginary components of the impedance become nearly indistinguishable, making any analysis much more difficult.  Using the expression for the admittance of the barrier and complex relative permittivity in \eqref{eq:Y_B_complex_perm} and \eqref{eq:complex_perm}, respectively, the low frequency approximations for the inner and outer contact impedances (\eqref{eq:Zi_lo} and \eqref{eq:Zo_lo}, respectively) can be written as
\begin{subequations}
	\begin{equation}
		Z_i^{\brk{lo}} \var{\omega} \approx \frac{1}{8 \pi \sigmach} + \frac{\tan{\var{\delta_\epsilon}}}{\omega \Delta_\epsilon \var{\omega} C_B^{\prime} \var{\omega} \pi r_i^2} \\
		+ \frac{1}{\imj \omega \Delta_\epsilon \var{\omega} C_B^{\prime} \var{\omega} \pi r_i^2}
	\end{equation}
	\begin{equation}
		Z_o^{\brk{lo}} \var{\omega} \approx \frac{1}{8 \pi \sigmach} \brk[s]*{ \frac{4 r_x^4}{\brk*{r_x^2 - r_o^2}^2} \ln{\brk*{\frac{r_x}{r_o}}} - \frac{2 r_x^2}{r_x^2 - r_o^2} - 1 } \\
		+ \frac{\tan{\var{\delta_\epsilon}}}{\omega \Delta_\epsilon \var{\omega} C_B^{\prime} \var{\omega} \pi \brk*{r_x^2 - r_o^2}} \\
		+ \frac{1}{\imj \omega \Delta_\epsilon \var{\omega} C_B^{\prime} \var{\omega} \pi \brk*{r_x^2 - r_o^2}}
	\end{equation}
\end{subequations}
where  $C_B^{\prime} \var{\omega} = \Real{\brk[s]{C_B \var{\omega}}} = \epsilon_r^{\prime} \var{\omega} \epsilon_0 / d_B$ is the real component of the barrier capacitance and $\Delta_\epsilon \var{\omega} = 1 +\tan^2{\var{\delta_\epsilon}} \ge 1$.  The associated low frequency approximations of the equivalent series resistance and capacitance of \eqref{eq:RT_CT} for a capacitor with a finite outer contact can then be shown to be
\begin{subequations}
	\begin{equation}\label{eq:RT_lo}
		R_T^{\brk{lo}} \var{\omega} = \Real{\brk[s]*{Z_T^{\brk{lo}} \var{\omega}}} \approx \frac{1}{8 \pi \sigmach} + \frac{1}{2 \pi \sigmach} \ln{\brk*{\frac{r_o}{r_i}}} \\
		+ \frac{1}{8 \pi \sigmach} \brk[s]*{ \frac{4 r_x^4}{\brk*{r_x^2 - r_o^2}^2} \ln{\brk*{\frac{r_x}{r_o}}} - \frac{2 r_x^2}{r_x^2 - r_o^2} - 1 } \\
		+ \frac{\tan{\var{\delta_\epsilon}}}{\omega \Delta_\epsilon \var{\omega} C_T^{\brk{0}} \var{\omega}}
	\end{equation}
	\begin{equation}\label{eq:CT_lo}
		C_T^{\brk{lo}} \var{\omega} = - \frac{1}{\omega \Imag{\brk[s]*{Z_T^{\brk{lo}} \var{\omega}}}} \approx \Delta_\epsilon \var{\omega} C_T^{\brk{0}} \var{\omega}
	\end{equation}
	where
	\begin{equation}\label{eq:CT_0}
		C_T^{\brk{0}} \var{\omega} = C_B^{\prime} \var{\omega} \brk[s]*{ \frac{1}{\pi r_i^2} + \frac{1}{\pi \brk*{ r_x^2 -r_o^2 }} }^{-1}
	\end{equation}
\end{subequations}
is the low frequency approximation of the equivalent series capacitance of the capacitor that ignores the imaginary component of the complex relative permittivity (i.e., $\epsilon_r^{\prime \prime} = 0$ and therefore $\tan{\var{\delta_\epsilon}} = 0$ and $\Delta_\epsilon \var{\omega} = 1$).  Similar expressions can be found for a capacitor with an infinite outer contact (see Appendix~\ref{app:low_freq_RT_CT_inf}).

Examining \eqref{eq:RT_lo}, it can be immediately seen that any frequency dependent behavior observed in the equivalent series resistance at low frequency is a result of the imaginary component of the complex relative permittivity, which manifests in the final, ``capacitive'' term of \eqref{eq:RT_lo}.  For cases where the imaginary component of the complex relative permittivity is zero or much, much less than the corresponding real component (i.e., when $\tan{\var{\delta_\epsilon}} = 0$ or $\ll 1$), this ``capacitive'' term is negligible and the equivalent series resistance is dominated by the resistive losses in the channel underneath the inner contact, gap, and outer contact, leading to a constant or nearly constant value for $R_T \var{\omega} \approx \eval{R_T^{\brk{lo}} \var{\omega}}_{\epsilon_r^{\prime \prime} = 0}$ (see Figs.~\ref{fig:circular_cap_dielectric}a and \ref{fig:circular_cap_Cole_Cole}a).  When the imaginary component of the complex relative permittivity is appreciably non-zero (i.e., when $\tan{\var{\delta_\epsilon}} \gtrsim 1$), the final, ``capacitive'' term of \eqref{eq:RT_lo} will tend to dominate the equivalent series resistance at low frequencies.  In other words, the barrier presents a larger dissipative element within the capacitor than the channel.  And the equivalent series resistance will increase markedly as the frequency decreases (see Figs.~\ref{fig:circular_cap_dielectric}b, c and \ref{fig:circular_cap_Cole_Cole}b, c).  For the specific case of a conductive barrier ($\epsilon_r \var{\omega} = \epsilon_r^{\prime} - \imj \sigma_B / \omega \epsilon_0$, Fig.~\ref{fig:circular_cap_dielectric}c), the material loss tangent is equivalent to the inverse product of the frequency and the dielectric relaxation time ($\tau_D$) of the barrier material: $\tan{\var{\delta_\epsilon}} = 1 / \omega \tau_D$ with $\tau_D = \epsilon_r^{\prime} \epsilon_0 / \sigma_B$.  As a result, the equivalent series resistance will increase with decreasing frequency until a frequency given by $f_D = 1 ⁄ 2 \pi \tau_D$, below which the equivalent series resistance becomes effectively constant with a value of $R_T \var{\omega} \approx \tau_D / C_T^{\brk{0}} \var{\omega} = \frac{d_B}{\sigma_B} \brk[s]2{ \frac{1}{\pi r_i^2} + \frac{1}{\pi \brk*{r_x^2 - r_o^2}} }$.

Again considering cases where the imaginary component of the complex relative permittivity is zero or much, much less than the corresponding real component (i.e., when $\tan{\var{\delta_\epsilon}} = 0$ or $\ll 1$), the imaginary component in relation to the real component of the complex relative permittivity can essentially be ignored and the equivalent series capacitance is effectively given by \eqref{eq:CT_0}: $C_T \var{\omega} \approx C_T^{\brk{lo}} \var{\omega} \approx C_T^{\brk{0}} \var{\omega}$ as $\Delta_\epsilon \var{\omega} \approx 1$ (see Figs.~\ref{fig:circular_cap_dielectric}a, b and \ref{fig:circular_cap_Cole_Cole}).  When the imaginary component of the complex relative permittivity is appreciably non-zero (i.e., when $\tan{\var{\delta_\epsilon}} \gtrsim 1$), the equivalent series capacitance will be larger than this value: $C_T \var{\omega} \approx C_T^{\brk{lo}} \var{\omega} = \Delta_\epsilon \var{\omega} C_T^{\brk{0}} \var{\omega}$ as $\Delta_\epsilon \var{\omega} > 1$.  Specifically for the case of a conductive barrier ($\epsilon_r \var{\omega} = \epsilon_r^{\prime} - \imj \sigma_B / \omega \epsilon_0$, Fig.~\ref{fig:circular_cap_dielectric}c), the equivalent series capacitance effectively becomes $C_T \var{\omega} \approx C_T^{\brk{lo}} \var{\omega} \approx C_T^{\brk{0}} \var{\omega} / \omega \tau_D$ at frequencies below $f_D = 1 ⁄ 2 \pi \tau_D$, leading to a clear increase over $C_T^{\brk{0}} \var{\omega}$ as frequency decreases.

Excepting the constant and real relative permittivity (Fig.~\ref{fig:circular_cap_dielectric}a) and Cole-Cole dielectric response with $\alpha_{cc} = 1$ (Fig.~\ref{fig:circular_cap_Cole_Cole}a) cases, the loss tangent, $\tan{\var{\delta}}$, of the overall capacitor reduces to that of the barrier material, $\tan{\var{\delta_\epsilon}}$, at low frequencies.  As noted previously, for these cases, the dissipative elements introduced by the imaginary component of the complex relative permittivity of the barrier dominate the resistive losses associated with the channel at low frequencies.  This effectively makes the barrier the overall dominat element within the impedance, for both dissipative and non-dissipative elements, at low frequencies and subsequently within the loss tangent of the entire capacitor:
\begin{equation}
		\tan{\var{\delta}} \approx - R_T^{\brk{lo}} \var{\omega} / X_T^{\brk{lo}} \var{\omega} \\
		\approx - R_T^{\brk{lo}} \var{\omega} \bigg/ \brk[s]*{ - \frac{1}{ \omega C_T^{\brk{lo}} \var{\omega} } } \\
		\approx \brk[s]*{ \frac{\tan{\var{\delta_\epsilon}}}{ \omega \Delta_\epsilon \var{\omega} C_T^{\brk{0}} \var{\omega}} } \bigg/ \brk[s]*{ \frac{1}{\omega \Delta_\epsilon \var{\omega} C_T^{\brk{0}} \var{\omega}} } \\
		\approx \tan{\var{\delta_\epsilon}} \, .
\end{equation}
For the constant and real relative permittivity (Fig.~\ref{fig:circular_cap_dielectric}a) and Cole-Cole dielectric response with $\alpha_{cc} = 1$ (Fig.~\ref{fig:circular_cap_Cole_Cole}a) cases, the imaginary component of the complex relative permittivity of the barrier at low frequencies is zero or nearly so.  As a result, the loss tangent of the overall capacitor effectively corresponds to a ratio between the resistive aspects of the channel and capacitive aspects of the barrier, as opposed to a fundamental behavior of the barrier relative permittivity (see Appendix~\ref{app:ColeCole} for a more detailed examination of the Cole-Cole dielectric response).

For all presented cases, at higher frequencies, the imaginary component of the complex relative permittivity of the barrier is or approaches zero, resulting in the loss tangent of the overall capacitor corresponding to a ratio between the resistive aspects of the channel and capacitive aspects of the barrier.  Correspondingly, the loss tangent has a strong size dependence, as seen in Figs.~\ref{fig:circular_cap_dielectric} and \ref{fig:circular_cap_Cole_Cole}, due to the incongruent size dependence of the various resistive and capacitive elements of the impedance.  Additionally, in all cases, the loss tangent of the overall capacitor is equal to or higher than that of the barrier material, by many orders of magnitude in some cases.  Therefore, improperly attributing this ``device'' behavior to that of the material dramatically overestimates the dissipative properties of the material and underestimates the its true performance.

As a final note, the examples presented here are not exhaustive.  The number of possible dielectric responses and associated behaviors are boundless and not mutually exclusive.

\section{Experimental Validation}

\subsection{ScAlN MIM Capacitor}
\begin{figure}[!htb]
	\centerline{\includegraphics[width=7.0cm]{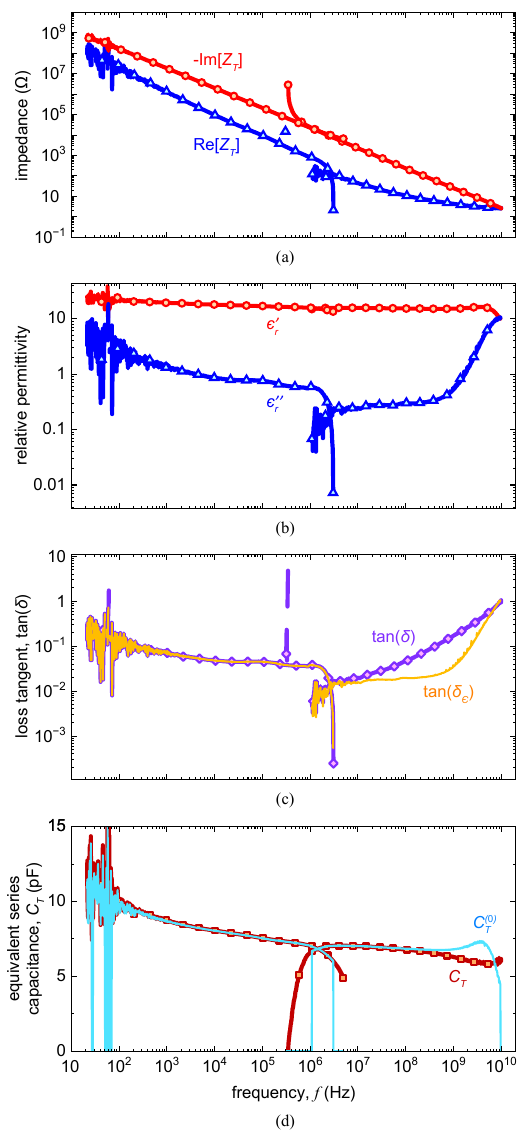}}
	\caption{(a) Measured real and imaginary impedance for a ScAlN MIM infinite outer contact circular capacitor ($r_i = 50$~{\textmu}m, $r_o = 70$~{\textmu}m).  (b) Extracted real ($\epsilon_r^{\prime}$) and imaginary ($\epsilon_r^{\prime \prime}$) relative permittivities ($\epsilon_r = \epsilon_r^{\prime} - j \epsilon_r^{\prime \prime}$).  (c) Measured loss tangent ($\tan{\var{\delta}}$) of the full capacitor and the barrier material ($\tan{\var{\delta_\epsilon}}$).  (d) Extracted equivalent series capacitance ($C_T \var{\omega}$ of \eqref{eq:CT}) with the approximate low-frequency equivalent series capacitance ($C_T^{\brk{0}} \var{\omega}$ of \eqref{eq:CT_0}).}
	\label{fig:ScAlN_MIM}
\end{figure}

Fig.~\ref{fig:ScAlN_MIM} shows the measured impedance (real and imaginary components), the extracted relative permittivity of the barrier (real and imaginary components, using \eqref{eq:Y_inner}, \eqref{eq:Y_outer_inf}, \eqref{eq:Z_gap}, and \eqref{eq:ZT_def}), the loss tangent of the full capacitor (evaluated from the measured impedance using \eqref{eq:tan_d}) and the material (evaluated from the extracted relative permittivity using \eqref{eq:tan_d_matl}), and the extracted equivalent series capacitance of \eqref{eq:CT} for a ScAlN MIM infinite outer contact circular capacitor with inner and outer radii of $r_i = 50$~{\textmu}m and $r_o = 70$~{\textmu}m, respectively.  The barrier consists of a 108~nm Sc$_{0.32}$Al$_{0.68}$N layer on a 47~nm AlN buffer ($d_B = 155$~nm).  The entire barrier sits upon a 52~nm NbN channel with a measured sheet resistance of 15.74~$\Omega / \square$, equal to a channel sheet conductance of $\sigmach = 0.0635$~S$\cdot\square$.  The gate contact consists of a Ni/Au (20/200~nm) stack.  An additional series resistance of 3~$\Omega$ due to contact and cable resistance was included in the total impedance during extraction procedures.  Measurements were made across a very broad frequency range using an impedance analyzer (2~ Hz – 50~MHz) and network analyzer (30~MHz – 10~GHz).  Near the edges of the respective analyzers' measurement ranges, the data becomes noisy due to the practical limitations of the analyzers.  This data is included within the various plots to demonstrate the degree of overlap between the analyzers and the limitations of the extraction.  The full details of the growth, fabrication, measurement, and analysis of these capacitors can be found in Ref. \cite{Gokhale2025}.

\begin{figure}[!htb]
	\centerline{\includegraphics[width=7.0cm]{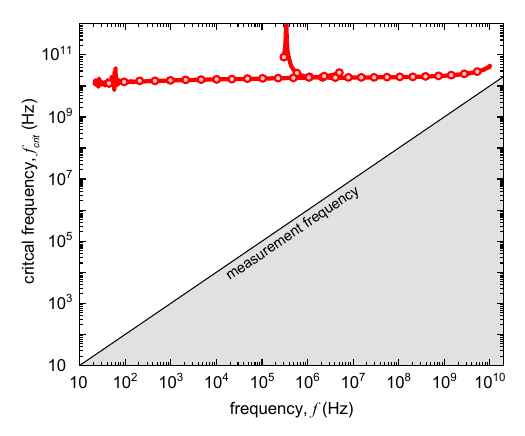}}
	\caption{Plot of critical frequency, $f_{crit}$ of \eqref{eq:fcrit}, versus measurement frequency for a ScAlN MIM infinite outer contact circular capacitor ($r_i = 50$~{\textmu}m, $r_o = 70$~{\textmu}m).  The critical frequency is greater than the measurement frequency across its entire range.}
	\label{fig:fcrit}
\end{figure}

Given that the NbN channel is metallic and the barrier layers are undoped, wide bandgap materials, the ScAlN MIM capacitor falls into the special metallic channel case, noted previously.  As a result, the model developed in this paper can be applied without restriction, and a point-by-point extraction of the relative permittivity (real and imaginary components) of the Sc$_{0.32}$Al$_{0.68}$N/AlN can be made (Fig.~\ref{fig:ScAlN_MIM}b).  The real component of the relative permittivity has a slight slope with respect to frequency, suggestive of a Cole-Cole or similar dielectric response, which is also reflected in the behavior of the equivalent series capacitance (Fig.~\ref{fig:ScAlN_MIM}d).  Assuming the AlN layer can be represented by a constant relative permittivity of $\epsilon_{r,AlN} = 10$ across all frequencies, the extracted real relative permittivity for the Sc$_{0.32}$Al$_{0.68}$N layer lies between approximately 50 at lower frequencies and 25 at higher frequencies.  Using \eqref{eq:fcrit}, the critical frequency ($f_{crit}$), signifying the maximum frequency up to which a low-frequency approximate $RC$-model would be applicable, is shown in Fig.~\ref{fig:fcrit}; across most of the measurement range, the critical frequency lies well above the measurement frequency.

\begin{figure}[!htb]
	\centerline{\includegraphics[width=7.0cm]{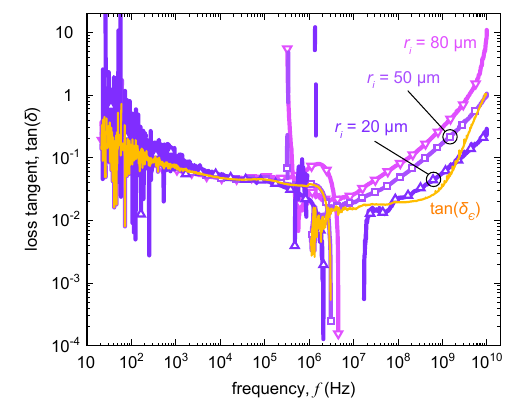}}
	\caption{Measured loss tangents,$\tan{\var{\delta}}$, of three different sized ScAlN MIM infinite outer contact circular capacitors ($r_i = 20$~{\textmu}m and $r_o = 28$~{\textmu}m, $r_i = 50$~{\textmu}m and $r_o = 70$~{\textmu}m, and $r_i = 80$~{\textmu}m and $r_o = 112$~{\textmu}m) and the extracted barrier material loss tangent, $\tan{\var{\delta_\epsilon}}$.  Note, the material loss tangent, $\tan{\var{\delta_\epsilon}}$, shown is extracted from the $r_i = 50$~{\textmu}m capacitor.}
	\label{fig:ScAlN_tand}
\end{figure}

Fig.~\ref{fig:ScAlN_MIM}d shows $C_T^{\brk{0}} \var{\omega}$ of \eqref{eq:CT_0} is essentially identical to the equivalent series capacitance extracted via \eqref{eq:CT}, owing to the imaginary component of the relative permittivity being much less than its real counterpart ($\epsilon_r^{\prime \prime} \var{\omega} \ll \epsilon_r^{\prime} \var{\omega}$, Fig.~\ref{fig:ScAlN_MIM}b).  Comparison of the measured loss tangent and material loss tangent (Fig.~\ref{fig:ScAlN_MIM}c) shows a clear discrepancy at higher frequencies and effective equivalence at lower frequencies, both as anticipated.  Fig.~\ref{fig:ScAlN_tand} shows the measured loss tangent for three differently sized capacitors with $r_i = 20$~{\textmu}m, 50~{\textmu}m, and 80~{\textmu}m and $r_o = 1.4 r_i$.  It can readily be seen from the figure that the loss tangent scales with size at higher frequencies (i.e., larger capacitors having larger loss tangent), as a direct result of the differing size dependencies of the various components of the capacitor; but regardless of size, the loss tangent reduces to that of the material loss tangent and overlap at lower frequencies, again all as anticipated.

\subsection{High-K Passivated GaN HEMT}
\begin{figure}[!htb]
	\centerline{\includegraphics[width=7.0cm]{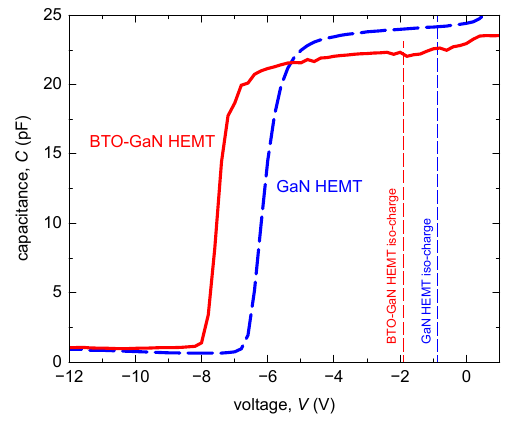}}
	\caption{Capacitance-voltage (CV) measurements of a GaN HEMT and a BTO-passivated GaN HEMT.  The iso-charge points of $-0.86$~V for the GaN HEMT and $-1.91$~V for the BTO-GaN HEMT are indicated on the plot.}
	\label{fig:HEMT_CV}
\end{figure}

Fig.~\ref{fig:HEMT_CV} shows the capacitance-voltage (CV) measurements of a GaN HEMT and a GaN HEMT passivated with MBE-grown epitaxial barium titanate (BTO) infinite outer contact circular capacitor with inner and outer radii of $r_i = 50$~{\textmu}m and $r_o = 70$~{\textmu}m.  The layer structure of the GaN HEMT, from the SiC substrate to surface, consists of a 100~nm AlN buffer layer, a 1~{\textmu}m GaN layer (i.e., the channel), and a 29~nm Al$_{0.25}$Ga$_{0.75}$N layer (i.e., the barrier).  For the BTO-GaN HEMT, an additional 40~nm BTO layer on a 1~nm/2~nm titanium oxide (TiO$_2$)/strontium titanate (STO) interfacial bi-layer was subsequently grown and adds to the ``barrier'' of the BTO-GaN HEMT.  The electron sheet concentration and mobility in the ungated GaN HEMT (BTO-GaN HEMT) channel is $1.47 \times 10^{13}$~cm$^{-2}$ ($1.49 \times 10^{13}$~cm$^{-2}$) and 674~cm$^2$/V$\cdot$s (746~cm$^2$/V$\cdot$s), respectively, giving a channel sheet conductance of $\sigmach = 1.59$~mS$\cdot \square$ (1.78~mS$\cdot \square$).

\begin{figure}[!htb]
	\centerline{\includegraphics[width=7.0cm]{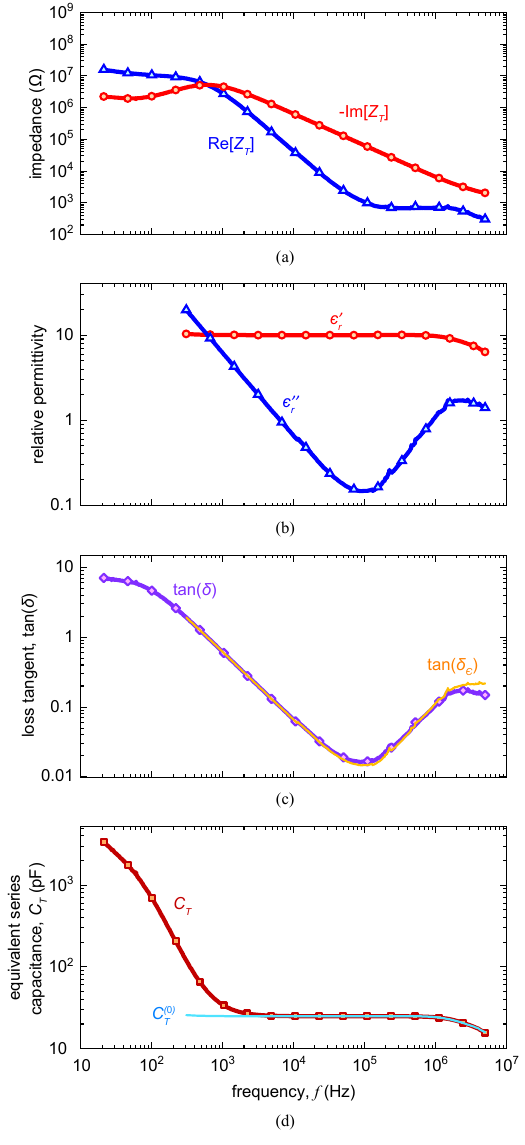}}
	\caption{(a) Measured real and imaginary impedance for a GaN HEMT infinite outer contact circular capacitor ($r_i = 50$~{\textmu}m, $r_o = 70$~{\textmu}m).  (b) Extracted real ($\epsilon_r^{\prime}$) and imaginary ($\epsilon_r^{\prime \prime}$) relative permittivities ($\epsilon_r = \epsilon_r^{\prime} - \imj \epsilon_r^{\prime \prime}$).  (c) Measured loss tangent ($\tan{\var{\delta}}$) of the full capacitor and the barrier material ($\tan{\var{\delta_\epsilon}}$).  (d) Extracted equivalent series capacitance ($C_T \var{\omega}$ of \eqref{eq:CT}) with the approximate low-frequency equivalent series capacitance ($C_T^{\brk{0}} \var{\omega}$ of \eqref{eq:CT_0}).}
	\label{fig:GaN_HEMT}
\end{figure}

Fig.~\ref{fig:GaN_HEMT} shows the measured impedance (real and imaginary components), the extracted relative permittivity of the full structure (real and imaginary components, using \eqref{eq:Y_inner}, \eqref{eq:Y_outer_inf}, \eqref{eq:Z_gap}, and \eqref{eq:ZT_def}), the loss tangent of the full capacitor (evaluated from the measured impedance using \eqref{eq:tan_d}) and the material (evaluated from the extracted relative permittivity using \eqref{eq:tan_d_matl}), and the extracted equivalent series capacitance of \eqref{eq:CT} for the GaN HEMT of Fig.~\ref{fig:HEMT_CV}, at its iso-charge bias point.  The iso-charge bias points are the relative bias conditions on each structure such that the charge in the HEMT channel is equal, resulting in the electronic state of the HEMT layers in both structures being identical (e.g., identical quantum capacitances).  For the devices presented here, the iso-charge bias points were $-0.86$~V and $-1.91$~V for a chosen channel concentration of $1 \times 10^{13}$~cm$^{-2}$ in both the GaN HEMT and BTO-GaN HEMT, respectively.  The measurements presented here were made using an impedance analyzer (20~Hz – 120~MHz).

The analysis of structures like a HEMT, where the channel conductance is bias dependent, requires more consideration than structures like a MIM capacitor, where the channel conductance is bias independent, though the analysis is still fairly straightforward.  In short, the three components of the full capacitor, the inner contact capacitor, the outer contact capacitor, and the gap in-between, can and will generally have different channel conductances.  However, assuming the mobility of the carriers in the channel remains effectively constant with respect to carrier density, a valid assumption for relatively minor changes in carrier density, the conductances in each component can be estimated.

The channel charge density in the gap between contacts can be assumed equivalent to that measured from an ungated Hall measurement, and therefore having a channel sheet conductance given by the results of the Hall measurement.  The channel charge density under the outer contact can be assumed equivalent to the zero-bias charge density calculated from a CV measurement on the capacitor, assuming the outer contact presents a lower impedance relative to the inner contact due to its larger area.  The associated channel conductance is calculated using that zero-bias charge density and the assumed unchanged mobility from the ungated Hall measurement.  The channel charge density under the inner contact is the charge density at the iso-charge bias point, with an associated channel conductance calculated using that charge density with, again, the assumed unchanged mobility of the ungated Hall measurement.  With the estimated channel conductances for the three components of the overall capacitor, the various dielectric and electronic properties of the capacitor can be extracted using \eqref{eq:Y_inner}, \eqref{eq:Y_outer_inf}, \eqref{eq:Z_gap}, and \eqref{eq:ZT_def}, some of which are shown in Fig.~\ref{fig:GaN_HEMT}.  The full details of the growth, fabrication, measurement, and analysis of these capacitors can be found in Ref. \cite{Jin2025}.

Comparison of Fig.~\ref{fig:GaN_HEMT}a to the impedance behavior shown in Fig.~\ref{fig:circular_cap_dielectric}c suggests that the AlGaN barrier of the GaN HEMT is leaky (i.e., conductive).  This is further supported by the low frequency behavior of the extracted imaginary component of the relative permittivity, loss tangent, and equivalent series capacitance shown in Figs.~\ref{fig:GaN_HEMT}b, c, and d, respectively, all which exhibit an inversely proportional dependence on frequency, similar to that seen for conductive barriers (e.g., $\epsilon_r^{\prime\prime} \propto \sigma_B / \omega \epsilon_0$, Fig.~\ref{fig:circular_cap_dielectric}c).  DC current-voltage measurements agree with this assessment.  The conductance of the entire capacitor is approximately $1 \times 10^{-8}$~S around the iso-charge bias point.  As noted previously, excessive barrier conductivity can invalidate the model developed in this paper for non-metallic channels; however, assigning the estimated measured DC conductance solely to the barrier gives a worst-case, approximate barrier conductance of $G_B \approx 1.27$~mS/cm$^2$, which although high, still easily satisfies the condition that the barrier be nearly insulating ($G_B \lambda^2 \ll \sigmach$ with $\lambda = r_i$) for a circular capacitor with an inner contact radius of $r_i = 50$~{\textmu}m, validating the application of the model developed in this paper under the special case of a (nearly) perfectly insulating barrier.  Without any assumption beyond this, a point-by-point extraction of the relative permittivity (real and imaginary components) and related properties of the barrier can be made (Fig.~\ref{fig:GaN_HEMT}).

\begin{figure}[!htb]
	\centerline{\includegraphics[width=7.0cm]{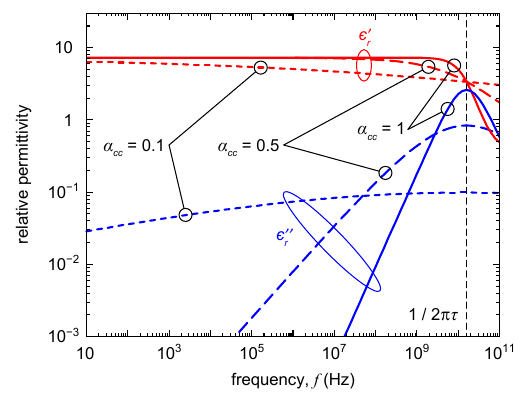}}
	\caption{Real and imaginary components of the relative permittivity for the Cole-Cole dielectric responses in Fig.~\ref{fig:circular_cap_Cole_Cole}.}
	\label{fig:Cole_Cole_perm}
\end{figure}

The behavior of the real and imaginary components of the relative permittivity at higher frequencies seen in Fig.~\ref{fig:GaN_HEMT}b is suggestive of a Cole-Cole-like dielectric response.  Fig.~\ref{fig:Cole_Cole_perm} shows the behavior of the real and imaginary components of the relative permittivity for the Cole-Cole dielectric responses shown in Fig.~\ref{fig:circular_cap_Cole_Cole}.
As can be seen from the figure, as the measurement frequency approaches the frequency associated with the characteristic time constant of the dielectric response, the real component of the relative permittivity drops from its constant low frequency value and the imaginary component rises from its near zero low frequency value, which is observed in Fig.~\ref{fig:GaN_HEMT}b.  As a result, it can be concluded that a composite dielectric behavior is being observed for the ``barrier'' (i.e., the AlGaN material layer and possibly AlGaN-GaN interface) of the GaN HEMT in Fig.~\ref{fig:GaN_HEMT} that consists of a combination of conductive leakage through the barrier, dominate at lower frequencies, and a Cole-Cole-like dielectric response, dominate at higher frequencies.  This conclusion is further supported by the behavior of the loss tangent in Fig.~\ref{fig:GaN_HEMT}c where the loss tangent increases at lower frequencies, associated with conduction through the barrier, and the loss tangent and material loss tangent are equivalent and increase at higher frequencies, suggesting the behavior is associated with the ``barrier'' material and not due to size effects.

\begin{figure}[!htb]
	\centerline{\includegraphics[width=7.0cm]{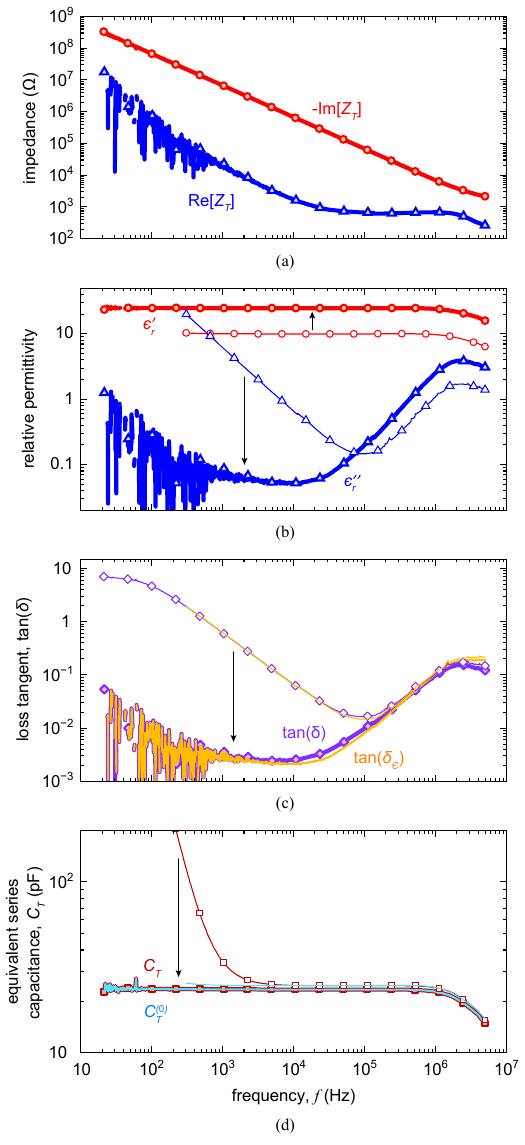}}
	\caption{(a) Measured real and imaginary impedance for a BTO-GaN HEMT infinite outer contact circular capacitor ($r_i = 50$~{\textmu}m, $r_o = 70$~{\textmu}m).  (b) Extracted real ($\epsilon_r^{\prime}$) and imaginary ($\epsilon_r^{\prime \prime}$) relative permittivities ($\epsilon_r = \epsilon_r^{\prime} - \imj \epsilon_r^{\prime \prime}$).  (c) Measured loss tangent ($\tan{\var{\delta}}$) of the full capacitor and the barrier material ($\tan{\var{\delta_\epsilon}}$).  (d) Extracted equivalent series capacitance ($C_T \var{\omega}$ of \eqref{eq:CT}) with the approximate low-frequency equivalent series capacitance ($C_T^{\brk{0}} \var{\omega}$ of \eqref{eq:CT_0}).  The thin lines in parts (b), (c), and (d) correspond to values from the GaN HEMT presented in Fig.~\ref{fig:GaN_HEMT}, showing the relative change in each parameter.}
	\label{fig:BTO_HEMT}
\end{figure}

Fig.~\ref{fig:BTO_HEMT} shows the measured impedance (real and imaginary components), the extracted relative permittivity of the full structure (real and imaginary components), the loss tangent of the full capacitor (evaluated from the measured impedance) and the material (evaluated from the extracted relative permittivity), and the extracted equivalent series capacitance for the BTO-passivated GaN HEMT of Fig.~\ref{fig:HEMT_CV}, at its iso-charge bias point.  For reference, some parameters from the GaN HEMT of Fig.~\ref{fig:GaN_HEMT} are included within the plots of Fig.~\ref{fig:BTO_HEMT}.

The differences between the BTO-passivated GaN HEMT in Fig.~\ref{fig:BTO_HEMT} and the GaN HEMT in Fig.~\ref{fig:GaN_HEMT} are immediately noticeable.  The addition of the BTO layer dramatically reduces the leakage current in the capacitor.  This directly leads to a reduced imaginary component of the relative permittivity (Fig.~\ref{fig:BTO_HEMT}b), a reduced loss tangent at low frequencies (Fig.~\ref{fig:BTO_HEMT}c), and a more conventional equivalent series capacitance (Fig.~\ref{fig:BTO_HEMT}d).  The BTO being a high-$\kappa$ oxide leads to an overall increase in the real component of the relative permittivity of the ``barrier'' (i.e., BTO-TiO$_2$-STO-AlGaN material stack), as seen in Fig.~\ref{fig:BTO_HEMT}b.  The capacitor still retains the Cole-Cole-like behavior seen in the GaN HEMT at high frequencies, suggesting whatever is driving that behavior is present within the GaN HEMT and is not tempered by the added high-$\kappa$ BTO layer.  From the measurements, the real component of the relative permittivity of the BTO-STO-TiO$_2$ oxide stack is estimated at approximately 340.

\section{Conclusions}
In this paper, a method for modeling the full steady-state and small-signal behavior of practical capacitive structures (``capacitors'') that can be used to understand and evaluate experimental results, extract various electronic and dielectric properties of the constituent materials, and to predict possible behavior of new structures and materials has been presented.  The archetypical behavior of a number of exemplar dielectric responses, including charge conduction through the capacitor (i.e., ``leakage''), was presented within a conventional circular capacitor layout.  Finally, the model and methodology presented in this paper was validated against experiment by examining the behaviors of a ScAlN MIM capacitor and a conventional and BTO-passivated GaN HEMT.

\section{Disclaimer}
The views, opinions and/or findings expressed are those of the authors and should not be interpreted as representing the official views or policies of the Department of Defense or the U.S. Government.

\clearpage

\appendix
\part*{\scshape Appendices}

\section{Constant Current in the Channel}\label{app:const_channel_current}
The total steady-state lateral current in the channel, $\vec{I}_{C\!H} \var{\vec{r}}$, in terms of the associated current sheet density, $\vec{K}_{C\!H} \var{\vec{r}}$, is
\begin{subequations}
	\begin{equation}
		\vec{I}_{C\!H} \var{\vec{r}} = 2 \pi r \vec{K}_{C\!H} \var{\vec{r}}
	\end{equation}
	and
	\begin{equation}
		\vec{I}_{C\!H} \var{\vec{r}} \approx W \vec{K}_{C\!H} \var{\vec{r}}
	\end{equation}
\end{subequations}
for circular and high aspect-ratio rectilinear (i.e., $W / L \gg 1$) geometries, respectively.  Applying \eqref{eq:char_eq_steady_state_Perfect}, which requires the divergence of the total steady-state lateral current sheet density in the channel to equal zero, using these expressions results in
\begin{subequations}\label{eq:div_KCH_zero_proof}
	\begin{equation}
			\nabla_r \cdot \vec{K}_{C\!H} \var{\vec{r}} = \nabla_r \cdot \brk*{ \frac{\vec{I}_{C\!H} \var{\vec{r}}}{2 \pi r} } \\
			= \frac{1}{r} \frac{\partial}{\partial r} \brk*{ r \frac{\vec{I}_{C\!H} \var{\vec{r}}}{2 \pi r} } \\
			= \frac{1}{2 \pi r} \frac{\partial}{\partial r} \vec{I}_{C\!H} \var{\vec{r}} = 0
	\end{equation}
	and
	\begin{equation}
			\nabla_r \cdot \vec{K}_{C\!H} \var{\vec{r}} = \nabla_r \cdot \brk*{ \frac{\vec{I}_{C\!H} \var{\vec{r}}}{W} } \\
			\approx \frac{\partial}{\partial x} \brk*{ \frac{\vec{I}_{C\!H} \var{\vec{r}}}{W} } \\
			\approx \frac{1}{W} \frac{\partial}{\partial x} \vec{I}_{C\!H} \var{\vec{r}} = 0
	\end{equation}
\end{subequations}
for circular and high aspect-ratio rectilinear geometries, respectively.  Therefore, as demonstrated in \eqref{eq:div_KCH_zero_proof}, regardless of geometry, \eqref{eq:char_eq_steady_state_Perfect} effectively requires that the total steady-state lateral current to be constant with respect to position in the lateral plane.

\section{High Aspect Ratio Rectilinear Capacitor}\label{app:rectilinear_cap}
\begin{figure}[!htb]
	\centerline{\includegraphics[width=7.0cm]{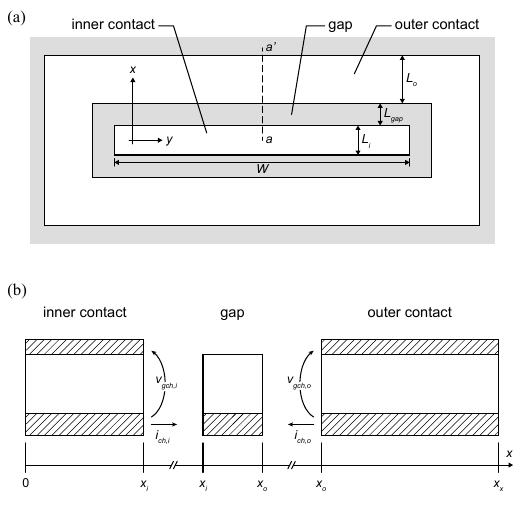}}
	\caption{(a) A rectilinear capacitor.  (b) Cross-sections of the constituent capacitor components along the cutline $a-a'$: $x_i = L_i ⁄ 2$, $x_o = x_i + L_{gap} = L_i ⁄ 2 + L_{gap}$, $x_x = x_o + L_o = L_i ⁄ 2 + L_{gap} + L_o$.}
	\label{fig:rectilinear_cap}
\end{figure}

Fig.~\ref{fig:rectilinear_cap} shows the structure of a generic capacitor of rectilinear geometry, consisting of an inner contact, an outer contact, and a gap between.
The inner contact has a length of $L_i$ and width of $W$.  It is assumed that the aspect ratio of the capacitor given by $W ⁄ L_i$  is very high, such that the capacitor can be assumed effectively uniform in the width direction ($y$-direction) and any non-uniformities introduced at the ends of the capacitor are negligible.  The length of the gap and outer contact are $L_{gap}$ and $L_o$, respectively.  For this analysis, it is assumed that the steady-state conductance through the barrier is sufficiently low, but not zero, such that the insulating barrier approximation may be used and, therefore, the channel sheet conductance and barrier admittance are effectively constant with respect to position.  For such a case with rectilinear geometry, \eqref{eq:char_eq_small_sig} becomes
\begin{equation}
	\frac{\partial^2}{\partial \chi^2} v_{gch} \var{ \chi,t } - \frac{Y_B \var{\omega} \lambda^2}{\sigmach} v_{gch} \var{\chi,t} = 0
\end{equation}
where $\chi = x / \lambda$ is the normalized lateral position.

Following a similar procedure as for the circular capacitor, the admittance of the inner contact capacitor is
\begin{equation}
		Y_i \var{\omega} = 2 \sigmach \frac{W}{L_i} \beta_i \tanh{\brk*{\frac{\beta_i}{2}}} \\
		= Y_B \var{\omega} L_i W \frac{2}{\beta_i} \tanh{\brk*{\frac{\beta_i}{2}}}
\end{equation}
with
\begin{equation}
	\beta^2 \var{\omega, x} = \frac{Y_B \var{\omega} \lambda^2}{\sigmach} \chi^2 = \frac{Y_B \var{\omega}}{\sigmach} x^2 = \mathscr{k}^2 \var{\omega} x^2
\end{equation}
being an analogue to $\alpha$ from \eqref{eq:alpha_def} in rectilinear coordinates.  With $\lambda = L_i$, $\beta_i = \sqrt{ Y_B \var{\omega} / \sigmach } L_i = \mathscr{k} L_i$.  The corresponding impedance is $Z_i \var{\omega} = 1 / Y_i \var{\omega}$.  The admittance of the outer contact capacitor is
\begin{equation}\label{eq:Y_outer_rect}
		Y_o \var{\omega} = 2 \sigmach \frac{W}{L_o} \beta_o \tanh{\brk*{\beta_o}} \\
		= Y_B \var{\omega} L_o W \frac{2}{\beta_o} \tanh{\brk*{\beta_o}}
\end{equation}
with $\lambda = L_o$ and $\beta_o = \sqrt{ Y_B \var{\omega} / \sigmach } L_o = \mathscr{k} L_o$ and an associated impedance of $Z_o \var{\omega} = 1 / Y_o \var{\omega}$.  Equation \eqref{eq:Y_outer_rect} is for a finite outer contact of length $L_o$.  The admittance for an outer contact capacitor with an infinite contact ($L_o \rightarrow \infty$) is
\begin{equation}
	Y_o^{\brk{\infty}} \var{\omega} = 2 W \sqrt{ Y_B \var{\omega} \sigmach }
\end{equation}
given $\tanh{\var{\beta}} \approx 1$ for $\beta \gg 1$.

As with the circular capacitors, for rectilinear capacitors where the outer contact is in direct electrical contact with the channel (e.g., the outer contact is ohmic), the impedance of the outer contact simply reduces to the contact resistance of that contact:
\begin{equation}
	Z_o \var{\omega} = \frac{1}{2} \frac{R_C}{W} \approx \frac{1}{2 \sigmach} \frac{L_T}{W}
\end{equation}
where $R_C \approx L_T ⁄ \sigmach$ is the contact resistance ($\Omega\cdot$mm) and $L_T$ is the transfer length of the contact and assuming that the width of the outer contact is much greater than the transfer length, $L_o \gg L_T$ \cite{Harrison1980}.  Note, the factor of $1/2$ comes from there being effectively two outer contacts in parallel.

The impedance of the gap between the inner and outer contacts is simply the resistance of the gap section of the channel between the contacts, again with the additional factor of $1/2$ due to the presence of two gaps in parallel:
\begin{equation}
	Z_{gap} \var{\omega} = \frac{1}{2 \sigmach} \frac{L_{gap}}{W} \, .
\end{equation}

The low frequency approximations of the various impedances are found using the approximate form of $\tanh{\var{\beta}}$ for values of $\beta \ll 1$ \cite{Zwillinger2003_6_7}:
\begin{equation}
	\tanh{\var{\beta}} \approx \beta \brk*{ 1 - \frac{\beta^2}{3} } \, .
\end{equation}
The resulting low frequency approximations for the inner contact capacitor and finite outer contact capacitor impedances are
\begin{equation}
	Z_i^{\brk{lo}} \var{\omega} \approx \frac{1}{Y_B \var{\omega} L_i W} + \frac{1}{12} \frac{1}{\sigmach} \frac{L_i}{W}
\end{equation}
and
\begin{equation}
	Z_o^{\brk{lo}} \var{\omega} \approx \frac{1}{2} \frac{1}{Y_B \var{\omega} L_o W} + \frac{1}{6} \frac{1}{\sigmach} \frac{L_o}{W} \, .
\end{equation}
In both cases, just as in the circular capacitor, the impedances are simply the series connection of the lumped element representations of the impedance of the inner or outer contact barrier and the spreading resistance of the associated conductive channel beneath \cite{Phillips1962}.  The impedance for the infinite outer contact capacitor remains unchanged.

The high frequency approximation for the inner contact, finite outer, and infinite outer capacitors are all identical and given by:
\begin{equation}
	Z_m^{\brk{hi}} \var{\omega} \approx \frac{1}{2W} \sqrt{ \frac{1}{Y_B \var{\omega} \sigmach} }
\end{equation}
with $m = i$ for the inner contact capacitor and $m = o$ for the outer contact capacitor.

The $RC$-like time constants for the inner and outer contact capacitors are
\begin{equation}
	\frac{Y_B \var{\omega} L_i W}{12 \sigmach W / L_i} \approx \frac{\imj \omega C_B \var{\omega} L_i^2}{12 \sigmach} = \imj \omega \tau_i \var{\omega}
\end{equation}
where $\tau_i \var{\omega} = C_B \var{\omega} L_i^2 / 12 \sigmach$, and 
\begin{equation}
	\frac{2 Y_B \var{\omega} L_o W}{6 \sigmach W / L_o} \approx \frac{\imj \omega C_B \var{\omega} L_o^2}{3 \sigmach} = \imj \omega \tau_o \var{\omega}
\end{equation}
where $\tau_o \var{\omega} = C_B \var{\omega} L_o^2 / 3 \sigmach$.  Assuming the outer contact is larger than or of similar length to the inner contact ($ L_i \le L_o$), the time constant of the outer contact capacitor will tend to dictate the critical frequency:
\begin{equation}
	f_{crit} \equiv f_o = \frac{1}{2 \pi \tau_o \var{\omega}} = \frac{1}{2 \pi} \frac{3 \sigmach}{C_B \var{\omega} L_o^2} \, .
\end{equation}

\section{The Low Frequency Approximate Equivalent Series Resistance and Capacitance of a Capacitor with an Infinite Outer Contact}\label{app:low_freq_RT_CT_inf}
Using the expression for the admittance of the barrier and complex relative permittivity in \eqref{eq:Y_B_complex_perm} and \eqref{eq:complex_perm}, respectively, the low frequency approximations for the outer contact impedance for an infinite outer contact of \eqref{eq:Zo_lo_inf} can be written as
\begin{equation}
	Z_o^{\brk{\infty,lo}} \var{\omega} \approx \frac{1}{2 \pi \sigmach} \brk[s]*{ \frac{1}{2} \ln{\brk*{ \frac{1}{2} \frac{8 \pi \sigmach \sqrt{ \Delta_\epsilon \var{\omega} } }{\omega \Delta_\epsilon \var{\omega} C_B^{\prime} \var{\omega} \pi r_o^2} } }  \\
	- \gamma - \imj \brk*{ \frac{\pi}{2} - \tan^{-1}{\brk*{ \tan{\var{\delta_\epsilon}} }} } }
\end{equation}
where  $C_B^{\prime} \var{\omega} = \Real \brk[s]{ C_B \var{\omega} } = \epsilon_r^{\prime} \var{\omega} \epsilon_0 / d_B$  is the real component of the barrier capacitance and $\Delta_\epsilon \var{\omega} = 1 + \tan^2{\var{\delta_\epsilon}} \ge 1$.  The associated low frequency approximations of the equivalent series resistance and capacitance of \eqref{eq:RT_CT} for a capacitor with an infinite outer contact can then be shown to be
\begin{subequations}
	\begin{equation}
		\begin{split}
		R_T^{\brk{\infty,lo}} \var{\omega} &= \Real \brk[s]*{ Z_T^{\brk{\infty,lo}} \var{\omega} } \\
		&\approx \frac{1}{8 \pi \sigmach} + \frac{1}{2 \pi \sigmach} \ln{\brk*{ \frac{r_o}{r_i} }} 
		+ \frac{1}{2 \pi \sigmach} \brk[s]*{ \frac{1}{2} \ln{\brk*{ \frac{1}{2} \frac{8 \pi \sigmach \sqrt{ \Delta_\epsilon \var{\omega} } }{\omega \Delta_\epsilon \var{\omega} C_B^{\prime} \var{\omega} \pi r_o^2} }} - \gamma } 
		+ \frac{\tan{\var{\delta_{\epsilon}}}}{\omega \Delta_\epsilon \var{\omega} C_T^{\brk{\infty,0}} \var{\omega}}
		\end{split}
	\end{equation}
	\begin{equation}
		\frac{1}{C_T^{\brk{\infty,lo}} \var{\omega}} = - \omega \Imag \brk[s]*{ Z_T^{\brk{\infty,lo}} \var{\omega} } \\
		\approx \frac{1}{\Delta_\epsilon \var{\omega} C_T^{\brk{\infty,0}} \var{\omega}} + \omega \brk*{ \frac{\pi}{2} - \tan^{-1}{\brk*{ \tan{\var{\delta_\epsilon}} }} } \, .
	\end{equation}
\end{subequations}
At low frequencies, this further reduces to
\begin{equation}
	\frac{1}{C_T^{\brk{\infty,lo}} \var{\omega}} \approx \frac{1}{\Delta_\epsilon \var{\omega} C_T^{\brk{\infty,0}} \var{\omega}}
\end{equation}
where
\begin{equation}
	C_T^{\brk{\infty,0}} \var{\omega} = C_B^{\prime} \var{\omega} \pi r_i^2
\end{equation}
is the low frequency approximation of the equivalent series capacitance of the capacitor that ignores the imaginary component of the complex relative permittivity (i.e., $\epsilon_r^{\prime \prime} = 0$ and therefore $\tan{\var{\delta_\epsilon}} = 0$ and $\Delta_\epsilon \var{\omega} = 1 $).

\section{Cole-Cole Dielectric Response and Loss Tangent}\label{app:ColeCole}
For the case of a barrier material with a Cole-Cole dielectric response of \eqref{eq:Cole_Cole}, the real and imaginary components of the complex permittivity are respectively \cite{ColeCole1941}
\begin{subequations}
	\begin{equation}
		\epsilon_r^{\prime} \var{\omega} = \epsilon_{r,\infty} \\
		+ \brk*{ \epsilon_{r,s} - \epsilon_{r,\infty} }
		\times \frac{ 1 + \brk*{\omega \tau}^{\alpha_{cc}} \sin{\!\brk*{ \frac{\pi}{2} \brk*{1 - \alpha_{cc}} }} }{ 1 + 2 \brk*{\omega \tau}^{\alpha_{cc}} \sin{\!\brk*{ \frac{\pi}{2} \brk*{1 - \alpha_{cc}} }} + \brk*{\omega \tau}^{2 \alpha_{cc}} }
	\end{equation}
	\begin{equation}
		\epsilon_r^{\prime \prime} \var{\omega} = \brk*{ \epsilon_{r,s} - \epsilon_{r,\infty} } \\
		\times \frac{ 1 + \brk*{\omega \tau}^{\alpha_{cc}} \cos{\!\brk*{ \frac{\pi}{2} \brk*{1 - \alpha_{cc}} }} }{ 1 + 2 \brk*{\omega \tau}^{\alpha_{cc}} \sin{\!\brk*{ \frac{\pi}{2} \brk*{1 - \alpha_{cc}} }} + \brk*{\omega \tau}^{2 \alpha_{cc}} }  \, .
	\end{equation}
\end{subequations}
At lower frequencies, these reduce to approximately
\begin{subequations}
	\begin{equation}
		\epsilon_r^{\prime} \var{\omega} \approx \epsilon_{r,s}
	\end{equation}
	\begin{equation}
		\epsilon_r^{\prime \prime} \var{\omega} \approx \brk*{ \epsilon_{r,s} - \epsilon_{r,\infty} } \brk*{\omega\tau}^{\alpha_{cc}} \cos{\!\brk*{ \frac{\pi}{2} \brk*{1 - \alpha_{cc}} }} \, .
	\end{equation}
\end{subequations}
The associated loss tangent of the material is
\begin{equation}\label{eq:Cole_Cole_tan_d}
		\tan{\var{\delta_\epsilon}} = \frac{\epsilon_r^{\prime \prime} \var{\omega}}{\epsilon_r^{\prime} \var{\omega}} \\
		\approx \brk*{\omega\tau}^{\alpha_{cc}} \brk*{ 1 - \frac{\epsilon_{r,\infty}}{\epsilon_{r,s}} } \cos{\!\brk*{ \frac{\pi}{2} \brk*{1 - \alpha_{cc}} }} \, .
\end{equation}
Inserting \eqref{eq:Cole_Cole_tan_d} into the final term of \eqref{eq:RT_lo} results in
\begin{equation}
	\frac{\tan{\var{\delta_\epsilon}}}{\Delta_\epsilon \var{\omega}} \frac{1}{\omega C_T^{\brk{0}} \var{\omega}} \\
	\approx \frac{1}{\omega^{1-\alpha_{cc}}} \frac{\tau^{\alpha_{cc}}}{C_T^{\brk{0}} \var{\omega}} \brk*{ 1 - \frac{\epsilon_{r,\infty}}{\epsilon_{r,s}} } \cos{\!\brk*{ \frac{\pi}{2} \brk*{1 - \alpha_{cc}} }}
\end{equation}
where it has been assumed $\tan{\var{\delta_\epsilon}} \ll 1$, valid for $\omega \tau \ll 1$, resulting in $\Delta_\epsilon \var{\omega} = 1 - \tan^2{\var{\delta_\epsilon}} \approx 1$.  Given $\alpha_{cc} \le 1$, this term will have an inverse dependence upon frequency and tend to dominate the equivalent series resistance at low frequencies, except for the special case of $\alpha_{cc} = 1$ (i.e., the Debye dielectric response, Fig.~\ref{fig:circular_cap_Cole_Cole}a).  For a Debye-like dielectric response, barring any anomalous behavior in the real component of the complex relative permittivity, this term becomes a constant, one that is relatively smaller than the remainder of the expressions in \eqref{eq:RT_lo}.  As a result, the loss tangent of the capacitor in this case is essentially equivalent to that of a material with an ``ideal'' relative permittivity (i.e., real and constant, Fig.~\ref{fig:circular_cap_dielectric}a), and the loss tangent does not reduce to the loss tangent of the material, even at low frequencies.

\clearpage

\section*{Acknowledgment}
This work was funded by the Defense Advanced Research Projects Agency, Microsystem Technology Office (DARPA MTO) under the COmpact Front-end Filters at the ElEment-level (COFFEE) program, and the NRL Base Program. We are grateful to Dr. Benjamin Griffin, Dr. Todd Bauer, and Dr. Zachary Fishman of DARPA for their programmatic and funding support.


\clearpage


\begin{thebibliography}{00}
	\bibliographystyle{IEEEtran}
	
	\bibitem{Lang1987} D. V. Lang, M. B. Panish, F. Capasso, J. Allam, R. A. Hamm, and A. M. Sergent, ``Measurement of heterojunction band offsets by admittance spectroscopy: InP/Ga$_{0.47}$In$_{0.53}$As,'' \emph{Appl. Phys. Lett.}, vol. 50, no. 12, pp. 736--738, Mar. 1987, 10.1063/1.98083.
	
	\bibitem{Macdonald1992} J. R. Macdonald, ``Impedance spectroscopy,'' \emph{Ann. Biomed. Eng.}, vol. 20, pp. 289--305, May 1992, 10.1007/BF02368532.
	
	\bibitem{Paszkiewicz2001} B. Paszkiewicz, ``Impedance spectroscopy analysis of AlGaN/GaN HFET structures,'' \emph{J. Cryst. Growth}, vol. 230, pp. 590--595, Sept. 2001, 10.1016/S0022-0248(01)01265-9.
	
	\bibitem{Shealy2008} J. R. Shealy and R. J. Brown, ``Frequency dispersion in capacitance-voltage characteristics of AlGaN/GaN heterostructures,'' \emph{Appl. Phys. Lett.}, vol. 92, no. 3, pp. 032101--3, Jan. 2008, 10.1063/1.2835708.
	
	\bibitem{Paszkiewicz2013} B. Paszkiewicz, M. Wosko, R. Paszkiewicz, and M. Tlaczala, ``Nondestructive method for evaluation of electrical parameters of AlGaN/GaN HEMT heterostructures,'' \emph{Phys. Status Solidi C}, vol. 10, no. 3, pp. 490--493, Jan. 2013, 10.1002/PSSC.201200709.
	
	\bibitem{Donahue2013} M. Donahue, B. L\"ubbers, M. Kittler, P. Mai, and A Schober, ``Impedance characterization of AlGaN/GaN Schottky diodes with metal contacts,'' \emph{Appl. Phys. Lett.}, vol. 102, no. 14, pp. 141607--4, Apr. 2013, 10.1063/1.4801643.
	
	\bibitem{Kohler2015} K. K\"ohler, W. Pletschen, B. Godejohann, S. M\"uller, H. P. Menner, and O. Ambacher, ``Admittance–voltage profiling of Al$_x$Ga$_{1-x}$N/GaN heterostructures: Frequency dependence of capacitance and conductance,'' \emph{J. Appl. Phys.}, vol. 118, no. 20, pp. 205702--8, Nov. 2015, 10.1063/1.4936125.
	
	\bibitem{Wu2017} T.-L. Wu, B. Bakeroot, H. Liang, N. Posthuma, S. You, N. Ronch, S. Stoffels, D. Marcon, and S. Decoutere, ``Analysis of the gate capacitance–voltage characteristics in p-GaN/AlGaN/GaN heterostructures,'' \emph{IEEE Electron Device Lett.}, vol. 38, no. 12, pp. 1696--1699, Dec. 2017, 10.1109/LED.2017.2768099.
	
	\bibitem{Casamento2022} J. Casamento, H. Lee, T. Maeda, V. Gund, K. Nomoto, L. van Deurzen, W. Turner, P. Fay, S. Mu, C. G. Van de Walle, A. Lai, H. Xing, and D. Jena, ``Epitaxial Sc$_x$Al$_{1-x}$N on GaN exhibits attractive high-K dielectric properties,'' \emph{Appl. Phys. Lett.}, vol. 120, pp. 152901--7, Apr. 2022, 10.1063/5.0075636.
	
	\bibitem{Giribaldi2022} G. Giribaldi, M. Pirro, M. Assylbekova, B. Herrera, G. Michetti, L. Colombo et al., ``X-Band multi-frequency 30\% compound SCALN microacustic resonators and filters for 5G-advanced and 6G applications,'' \emph{2022 Joint Conference of the European Frequency and Time Forum and IEEE International Frequency Control Symposium (EFTF/IFCS)}, Paris, France, 2022, pp. 1--4, 10.1109/EFTF/IFCS54560.2022.9850563.
	
	\bibitem{Wang2020} D. Wang, J. Zheng, P. Musavigharavi, W. Zhu, A. C. Foucher, S. E. Trolier-McKinstry, E. A. Stach, R. H. Olsson, ``Ferroelectric switching in sub-20 nm aluminum scandium nitride thin films,'' \emph{IEEE Electron Device Lett.}, vol. 41, no. 12, pp. 1774--1777, Dec. 2020, 10.1109/LED.2020.3034576.
	
	\bibitem{Pradhan2024} D. K. Pradhan, D. C. Moore, G. Kim, Y. He, P. Musavigharavi, K.-H. Kim, N. Sharma, Z. Han, X. Du, V. S. Puli, E. A. Stach, W. J. Kennedy, N. R. Glavin, R. H. Olsson III, D. Jariwala, ``A scalable ferroelectric non-volatile memory operating at 600 $^\circ$C,'' \emph{Nat. Electron.}, vol. 7, no. 5, pp. 348--355, May 2024, 10.1038/s41928-024-01148-6.
	
	\bibitem{Xia2019} Z. Xia, C. Wang, N. K. Kalarickal, S. Stemmer, and S. Rajan, ``Design of transistors using high-permittivity materials,'' \emph{IEEE Trans. on Electron Devices}, vol. 66, no. 2, pp. 896--900, Feb. 2019, 10.1109/TED.2018.2888834.
	
	\bibitem{Rahman2021} M. W. Rahman, N. K. Kalarickal, H. Lee, T. Razzak, S. Rajan, ``Integration of high permittivity BaTiO$_3$ with AlGaN/GaN for near-theoretical breakdown field kV-class transistors,'' \emph{Appl. Phys. Lett.}, vol. 119, no. 19, pp. 193501--5, Nov. 2021, 10.1063/5.0070665.
	
	\bibitem{Zwillinger2003_6_19} D. Zwillinger, ``Special Functions: Bessel Functions,'' in \emph{CRC Standard Mathematical Tables and Formulae}, 31$^{\text{st}}$ ed., New York, NY, USA: CRC Press, 2003, ch. 6, sec. 6.19.
	
	\bibitem{Willis1987} A. J. Willis and A. P. Botha, ``Investigation of Ring Structures for Metal-Semiconductor Contact Resistance Determination,'' \emph{Thin Film Solids}, vol. 146, no. 1, pp. 15--20, Jan. 1987, 10.1016/0040-6090(87)90335-X.
	
	\bibitem{Phillips1962} A. B. Phillips, ``Characteristics of Junction Transistors,'' in \emph{Transistor Engineering and Introduction to Integrated Semiconductor Circuits}, New York, NY, USA: McGraw-Hill Book Company, 1962, ch. 9, sec. 9-6.
	
	\bibitem{ColeCole1941} K. S. Cole and R. H. Cole, ``Dispersion and Absorption in Dielectrics I. Alternating Current Characteristics,'' \emph{J. Chem. Phys.}, vol. 8, pp. 341--351, Apr. 1941, 10.1063/1.1750906.
	
	\bibitem{Gokhale2025} V. J. Gokhale, J. G. Champlain, M. T. Hardy, J. L. Hart, A. C. Lang, S. Mukhopadhyay, J. A. Roussos, S. C. Mack, G. Giribaldi, L. Colombo, M. Rinaldi, B. P. Downey, ``Strongly Dispersive Dielectric Properties of High-ScN-Fraction ScAlN deposited by Molecular Beam Epitaxy,'' [Online], Feb. 2025. Available: \underline{https://arxiv.org/abs/2502.04539}
	
	\bibitem{Jin2025} E. N. Jin, V. J. Gokhale, J. G. Champlain, J. L. Hart, A. C. Lang, M. T. Hardy, N. Nepal, D. S. Katzer, B. P. Downey, and V. D. Wheeler, ``High Permittivity Epitaxial Barium Titanate Thin Films Integrated on Gallium Nitride-based Devices for Radio Frequency Electronics,'' \emph{unpublished}.
	
	\bibitem{Harrison1980} H. B. Harrison, ``Characterising Metal Semiconductor Ohmic Contacts,'' \emph{Proceedings of the IREE Aust.}, vol. 41, no. 3, pp. 95-100, Sept. 1980.
	
	\bibitem{Zwillinger2003_6_7} D. Zwillinger, ``Special Functions: Hyperbolic Functions,'' in \emph{CRC Standard Mathematical Tables and Formulae}, 31$^{\text{st}}$ ed., New York, NY, USA: CRC Press, 2003, ch. 6, sec. 6.7.
\end{thebibliography}
\end{document}